\documentstyle[prd,aps,preprint,eqsecnum,epsfig,floats]{revtex}
\def\pnul{\raise-.3ex\hbox{$\stackrel{\circ}{p}$}}\relax
\def\epsnul{\raise.1ex\hbox{$\stackrel{\circ}{\varepsilon}$}}\relax
\tighten
\begin{document}
\title{Frame Dependence of Spin-One Angular Conditions in Light Front 
Dynamics}

\author{Bernard L. G. Bakker$^1$ and Chueng-Ryong Ji$^2$}

\address{$^1$ Department of Physics and Astrophysics, Vrije Universiteit, De 
Boelelaan 1081, NL-1081~HV~Amsterdam, The Netherlands}
\address{$^2$ Department of Physics, North Carolina State University,
Raleigh, NC 27695-8202, USA}

\date{DRAFT: \today}

\maketitle

\begin{abstract}

We elaborate the frame dependence of the angular conditions for spin-1 
form factors. An extra angular condition is found in addition to the 
usual angular condition relating the four helicity amplitudes.
Investigating the frame-dependence of angular conditions, we find
that the extra angular condition is in general as complicated as
the usual one, although it becomes very simple in the
$q^+ = 0$ frame involving only two helicity amplitudes.
It is confirmed that the angular conditions are identical in
frames that are connected by kinematical transformations.
The high $Q^2$ behaviors of the physical form factors and 
the limiting behaviors in special reference frames are also discussed.

\end{abstract}
\pacs{ }
\section{Introduction}
\label{sec.I}
Bosons with spin-1 are ubiquitous in modern particle physics. In the
Standard Model the fundamental interactions are described by
gauge-bosons, such as the photon, $W^\pm$ and $Z$, and the gluon.
These particles are considered to be truly elementary, {\it i.e.} they occur
as quanta of local fields.

In hadron physics many vector mesons composed of a quark and an antiquark are 
found and understanding their structure is a challenging problem in
quantum chromodynamics (QCD), related to the mechanism of confinement and the 
detailed nature of the interactions between the constituents.  

Moreover, the deuteron is an interesting laboratory for
the application of QCD to nuclear physics. At large distances the
deuteron is evidently well described as a spin-1 composite of two
nucleon clusters with binding energy $\sim 2.2$ MeV, together with
small admixtures of $\Delta \Delta$ and virtual meson components.
However, at short distances, in the region where all six quarks overlap
within a distance $R\sim 1/Q$, one can show rigorously that the deuteron
state in QCD necessarily has ``fractional parentage" $1/9\, (np)$,
$4/45\, (\Delta\Delta)$ and $4/5$ "hidden color" (nonnuclear)
components\cite{harvey}. At any momentum scale, the deuteron cannot be
described solely in terms of conventional nuclear physics degrees of
freedom, but in principle any dynamical property of the deuteron is
modified by the presence of non-abelian "hidden color"
components\cite{BrodskyJi}. Alternatively one may describe the deuteron
structure in terms of uncolored degrees of freedom only, but then a
tower of excited nucleons and $\Delta$'s are
involved\cite{NaroBen,Vento}.

Although these spin-1 systems  ({\it e.g.} $W^\pm$, the
$\rho-$meson and the deuteron) do not seem to share a common internal
structure, universality of spin-1 systems\cite{BrodskyHiller}
severely constrains them. According to this universality, the
fundamental constraints on the magnetic and quadrupole moments of
hadronic and nuclear states imposed by the Compton-scattering
sum-rules\cite{DHG} and the behavior of the electromagnetic form
factors of composite spin-1 systems\cite{ArnoldCarlsonGross} at large
momentum transfer are the same as those of a corresponding elementary
particle of the same spin and charge.  At $Q^2=0$, the charge ($G_C
(Q^2)$), magnetic ($G_M (Q^2)$) and quadrupole ($G_Q (Q^2)$) form
factors define the charge $e$, the magnetic moment $\mu_1$ and the
quadrupole moment $Q_1$, respectively. In the limit of zero radius of
the bound states (or large binding energies), whether
confined or non-confined, the values of $\mu_1$ and $Q_1$ approach
the canonical values\cite{BrodskyHiller} of a spin-1 object with
mass $m$ and charge $e$
\begin{equation}
 \mu_1 = \frac{e}{m}, \quad Q_1 = - \frac{e}{m^2}.
 \label{eq.1.10}
\end{equation}
Universality requires
that the values obtained in Eq.~(\ref{eq.1.10}) must be the same as those
of the fundamental gauge bosons $W^\pm$ in the tree approximation to the 
standard model. 
Also, at large $Q^2$ (in the limit $Q \gg \sqrt{2m\Lambda_{QCD}}$), 
these form factors are required to 
approach the universal ratios given by\cite{BrodskyHiller}
\begin{equation}
 G_C(Q^2):G_M(Q^2):G_Q(Q^2) \rightarrow \left(1-\frac{Q^2}{6m^2}\right):2:-1,
 \label{eq.1.20}
\end{equation}
which were obtained in a light-cone frame with $q^+ =0$.
Eq.~(\ref{eq.1.20}) should hold at large momentum transfers in the case
of composite systems such as the $\rho$-meson and the deuteron, with
corrections of order $\Lambda_{QCD}/Q$ and $\Lambda_{QCD}/m$ according
to QCD. The ratios are the same as those predicted for the
electromagnetic couplings of the $W^\pm$ for all $Q^2$ in the standard
model at tree level.

Furthermore, there are constraints on the current matrix elements,
since there are only three form factors for the spin-1 systems. 
A constraint well-known from the literature\cite{CarlsonHillerHolt}
is the angular condition obtained by demanding rotational covariance 
for the current matrix elements given by
\begin{equation}
G^\mu_{h^\prime,h} = \langle p^\prime h^\prime |J^\mu|p h \rangle,
 \label{eq.1.30}
\end{equation}
where $|p h \rangle$ is an eigenstate with momentum $p$ and helicity $h$.
For example, in the Drell-Yan-West (DYW) frame and the frames that are 
connected to DYW by only kinematic transformations, the angular condition is 
given  as \cite{BrodskyHiller,CarlsonHillerHolt,Keister} %
\begin{equation}
 (1+ 2\eta)G^+_{++} + G^+_{+-} - \sqrt{8\eta} G^+_{+0} - G^+_{00} =0,
 \label{eq.1.40}
\end{equation}
where $\eta = Q^2/4m^2$. 
If the angular condition is not satisfied, an identical extraction
of form factors ($G_C,G_M,G_Q$) from the light-front current matrix
elements $G^+_{h^\prime h}$ is not attained. As a consequence, there
are indeed different extraction schemes for the spin-1 form factors in
the literature\cite{BrodskyHiller,GrachKondratyuk,Chungetal,FFS}.
As an example, $G_C,G_M$ and $G_Q$ can be given in terms 
of $G^+_{+0},G^+_{00}$ and $G^+_{+-}$ in the DYW frame $q^+=0$, $q_x = Q$,
and $q_y=0$ as follows\cite{BrodskyHiller};
\begin{eqnarray}
 G_C &=& \frac{1}{2p^+(2\eta+1)}\left[\frac{16}{3}\eta\frac{G^+_{+0}}
 {\sqrt{2\eta}}-\frac{2\eta-3}{3}G^+_{00}+\frac{2}{3}(2\eta-1)G^+_{+-}\right],
 \nonumber \\
 G_M &=& \frac{2}{2p^+(2\eta+1)}\left[(2\eta-1)\frac{G^+_{+0}}
 {\sqrt{2\eta}}+G^+_{00}-G^+_{+-}\right],
 \nonumber \\
 G_Q &=& \frac{1}{2p^+(2\eta+1)}\left[2\frac{G^+_{+0}}
 {\sqrt{2\eta}}-G^+_{00}+\frac{\eta+1}{\eta}G^+_{+-}\right].
 \label{eq.1.50}
\end{eqnarray}
However, other choices of the current matrix elements can be made
to express the right-hand-side of Eq.~(\ref{eq.1.50}) and the 
expression also depends on the reference frame. A few 
examples of other expressions on the right-hand-side of 
Eq.~(\ref{eq.1.50}) can be found in Ref.\cite{MeloFrederico}. 
The angular conditions are also useful in testing the validity of model
calculations. Especially, as stressed in the recent literature
\cite{CJ,Jaus,BH,Matthias,Griffin}, the zero-mode contribution is necessary 
to get the correct result for the form factors unless the good component of the 
current is used. Even if the good component of the current is used, 
it was noted that the zero-mode contribution is necessary 
for the calculation of spin-1 form factors\cite{Meloetal}. 
Such an observation of zero-mode necessity has been made by checking
the angular conditions and the degree of neccessity
can be quantified by examining the angular conditions.

As discussed above, the constraints from universality and the angular 
conditions are in principle very useful for model-building and a 
self-consistency-check of theoretical or phenomenological models
for spin-1 objects. However, these constraints do depend 
on the reference frame. 
For example, in the Breit frame where $q^+ \neq 0$, a less informative 
prediction of asymptotic form factors is made\cite{CarlsonGross}
instead of Eq.~(\ref{eq.1.20});
\begin{equation} 
 G_C(Q^2):G_Q(Q^2) \rightarrow \frac{Q^2}{6m^2}:1
 \label{eq.1.60}
\end{equation}
in the limit $Q \gg 2m$.
Thus, it is important to examine the frame dependence of the constraints 
that are useful for model-building and phenomenology.

In this work, we analyze the frame dependence of the angular conditions
for the spin-1 systems. Interestingly, besides the angular condition
given by Eq.~(\ref{eq.1.40}) we find another one.  Elaborating the
frame-dependence of these angular conditions in the generalized Breit
and target rest frames as well as the DYW frame, we confirm the
advantage of using the DYW frame in the calculation of exclusive
processes. The complexity of each angular condition in general depends
on the reference frame.  In the DYW frame, the extra angular condition
is particularly simple so that most theoretical models are expected to
satisfy it without any difficulty. We also substantiate that the
angular conditions are identical in reference frames that are connected
by kinematical transformations.  Such an investigation is also
important in analyzing the high $Q^2$ behavior of spin-1 form factors.
We confirm that the angular conditions are consistent with the high
$Q^2$ behavior predicted by perturbative QCD (PQCD) for the three
physical form factors\cite{BrodskyHiller,CarlsonHillerHolt,Keister}.

In the next Section (Section II), both the instant-form (IFD)
and the front-form (LFD) polarization vectors are presented 
in arbitrary frames. In Section III, we derive the relation between
the current operator and the form factors and starting from general grounds
obtain the most general angular conditions. We show that there are indeed
two angular conditions and discuss the reason why they should be regarded
as consistency conditions. In Section IV, we elaborate the details of
the angular conditions in the DYW, generalized Breit and target-rest frames.
In Section V, we discuss the large momentum transfer behavior of the form 
factors in each reference frame. We also consider the limiting 
behaviors of the form factors in approaching special Breit and 
target-rest frames. Conclusions follow in Section VI. In the Appendix,
explicit representations of front-form boost and helicity operators 
are summarized. 

\section{Polarization Vectors in Instant-Form and Light-Front Dynamics}
\label{sec.II}
\subsection{Polarization Vectors in Three Dimensions}
\label{sec.II.A}
For the polarization vectors we use the standard spherical tensors for spin-1
\cite{BS}
\begin{equation}
 \vec{e}\, (0)  =  (0,0,1),  \quad
 \vec{e}\, (\pm)  =  \mp \frac{1}{\surd 2} (1, \pm i, 0).
 \label{eq.2.10}
\end{equation}
Complex conjugation gives
\begin{equation}
 \vec{e}\, (h)^* = (-1)^h \, \vec{e}\, (-h).
 \label{eq.2.20}
\end{equation}
The orthogonality relation is
\begin{equation}
 \vec{e}\, (h) \cdot \vec{e}\, (h')^* = \delta_{h h'},
 \label{eq.2.30}
\end{equation}
which can also be written as
\begin{equation}
 \vec{e}\, (h) \cdot \vec{e}\, (-h') = (-1)^h \,\delta_{h h'} .
 \label{eq.2.40}
\end{equation}
The closure property can be written as
\begin{equation}
 \sum_h e_i\,(h) \; e_j\,(h)^* = \delta_{i j} .
 \label{eq.2.50}
\end{equation}

\subsection{Polarization Vectors in Four Dimensions}
\label{sec.II.B}
It is easy to extend these three polarization vectors to four vectors:
\begin{equation}
 \epsnul \,(h) = (0, \vec{e}\, (h)) .
 \label{eq.2.60}
\end{equation}
Then the orthogonality and closure properties are
\begin{equation}
 \epsnul (h) \cdot \epsnul(h')^* = -\delta_{h h'} ,
 \label{eq.2.70}
\end{equation}
and
\begin{equation}
 \sum_h \epsnul_\mu \,(h) \;
 \epsnul_\nu \,(h)^* =
 -\left(g_{\mu \nu} -\frac{\pnul_\mu \pnul_\nu}{m^2}\right),
 \label{eq.2.80}
\end{equation}
where $\pnul_\mu = (m,0,0,0)$.

\subsection{Polarization Vectors in a Specific Lorentz Frame}
\label{sec.II.C}
We now consider the polarization vectors given above as belonging to a
particle of mass $m$ in its rest frame. Then the definition Eq.~(\ref{eq.2.60})
reflects the transversality property
\begin{equation}
 p_\mu \, \varepsilon^\mu(p;h) = 0 .
 \label{eq.2.90}
\end{equation}

In order to extend this property to all Lorentz frames, we define the
polarization vectors in a specific frame by boosting the vectors
(Eq.~(\ref{eq.2.60})) to the specific frame. The vectors we obtain will
depend on the Lorentz transformation.  Later on, we shall be interested
in the kinematic operators only, the number of which is maximized in
LFD.  In IFD all boosts are dynamical, so we cannot impose the same
limitation. As the instant-form results are for the purpose of
illustration only, we shall not worry about this point, but just limit
ourselves to pure Lorentz transformations,
{\it i.e.} rotationless boosts.

\subsection{Front-Form Polarization Vectors}
\label{sec.II.D}
In the front form we need the kinematical front-form boosts. They are given
in Appendix~\ref{sec.A}.

We note that the LF components we use satisfy the following relations
\begin{equation}
 \vec{p}^{\, 2}_\perp = -2 p^r p^l , \quad 
 p \cdot q = p^+ q^- + p^- q^+ + p^r q^l + p^l q^r,
 \label{eq.2.100}
\end{equation}
where we use the spherical components of the three momentum vectors to simplify
the notation>. They are defined as follows
\begin{equation}
 p^r = -\frac{p_x + i p_y}{\surd 2}, \quad p^l = \frac{p_x - i p_y}{\surd 2}.
 \label{eq.2.110}
\end{equation}
Occasionally we use the notation $p^h$ with $h=+1, 0, -1$ for $p^r$, $p_z$, and
$p^l$, respectively. Then the usual relation for spherical tensors applies
\begin{equation}
 (p^h)^* = (-1)^h p^{-h}.
 \label{eq.2.120}
\end{equation}

The polarization vectors in the rest system, where the four momentum has
the LF components $(p^+, p^1, p^2, p^-) = (m/\surd 2, 0,0, m/\surd 2)$, are
\begin{equation}
 \epsnul_{\rm ff}(\pm) = (0, \mp 1/\surd2, -i/\surd2, 0) , \quad
 \epsnul_{\rm ff}(0) = (1/\surd 2, 0,0, -1/\surd 2).
 \label{eq.2.130}
\end{equation}
Upon application of the front-form boost Eq.~(\ref{eq.A.120}) we find the
polarization vectors
\begin{equation}
 \left.
 \begin{array}{c}
 \varepsilon_{\rm ff} (p^+,p^1,p^2 ;+) \\
 \varepsilon_{\rm ff} (p^+,p^1,p^2 ;0) \\
 \varepsilon_{\rm ff} (p^+,p^1,p^2 ;-)
 \end{array} 
 \right\} = \left\{
 \begin{array}{c}
 \left( 0, \frac{-1}{\surd 2}, \frac{-i}{\surd 2},
 \frac{p^r}{p^+} \right) \\
 \left( \frac{p^+}{m}, \frac{p^1}{m}, \frac{p^2}{m},
 \frac{\vec{p}^{\perp \, 2} - m^2}{2 m p^+} \right) . \\
 \left( 0, \frac{ 1}{\surd 2}, \frac{-i}{\surd 2},
 \frac{p^l}{p^+} \right)
 \end{array} \right.
 \label{eq.2.140}
\end{equation}
It is easy to check that these are mutually orthogonal, transverse, and
satisfy the closure property Eq.~(\ref{eq.2.80}) if one uses the front-form 
for the metric.

\section{Currents}
\label{sec.III}
For a spin-1 particle the current operator has the form
\begin{equation}
 J^\mu_{\alpha \beta} (p^\prime, p) =
 - g_{\alpha \beta} (p^\prime + p)^\mu \, F_1(q^2) 
 + (g^\mu_\beta q_\alpha - g^\mu_\alpha q_\beta) F_2(q^2)
 +\frac{q_\alpha q_\beta (p^\prime + p)^\mu}{2 m^2} F_3(q^2),
 \label{eq.3.10}
\end{equation}
where the momenta $p$ and $p^\prime$ are the momenta of the particle 
before and after absorption of a photon with momentum $q = p^\prime - p$.
The coefficient functions $F_i(Q^2)$ in Eq.~(\ref{eq.3.10}) are given by
the physical form factors,{\it i.e.}
\begin{eqnarray}
F_1 &=& G_C - \frac{2}{3}\eta G_Q \nonumber \\
F_2 &=& -G_M \nonumber \\
F_3 &=& \frac{1}{1+\eta}\left[ -G_C + G_M + (1+\frac{2}{3}\eta)G_Q\right].
 \label{eq.3.20}
\end{eqnarray}
A spin tensor $G$ is obtained by taking matrix elements with the polarization 
vectors, viz
\begin{equation}
 G^\mu_{h^\prime h} =
 \varepsilon^{*\,\alpha}(p^\prime; h^\prime) \, J^\mu_{\alpha \beta}
 \, \varepsilon^\beta(p; h).
 \label{eq.3.30}
\end{equation}
This form can be derived on very general grounds. First, we write down all
tensors of third rank that can be constructed using $g_{\alpha\beta}$,
$p^{\prime\mu}$, and $p^\mu$ alone. There are fourteen possible structures.
As the matrix elements are obtained by contracting with the polarization
vectors $\varepsilon^{*\,\alpha}(p^\prime; h^\prime)$ and 
$\varepsilon^\beta(p; h)$, the structures containing a factor 
$p^\prime_\alpha$ or $p_\beta$ do not contribute to the matrix element.
Therefore, only six remain and we write
\begin{equation}
 J^\mu_{\alpha\beta}(p',p) =
   f_1 \, g_{\alpha\beta} \, p^{\prime\mu}
 + f_2 \, g_{\alpha\beta} \, p^\mu
 + f_3 \, g^\mu_\alpha \, p^\prime_\beta
 + f_4 \, g^\mu_\beta p_\alpha
 + f_5 \, p^{\prime\mu} p_\alpha p^\prime_\beta
 + f_6 \, p^\mu p_\alpha p^\prime_\beta.
 \label{eq.3.40}
\end{equation}

Secondly, we require current conservation, which means
$q_\mu G^\mu_{h' h} (p', p) = 0$ for all $\mu$, $h'$, and $h$. 
Contracting with $q$ gives
\begin{equation}
 0 = (f_1 - f_2) g_{\alpha\beta} (m^2 - p' \cdot p)
 + f_3 q_\alpha  p^\prime_\beta + f_4 q_\beta p_\alpha
 + (f_5 - f_6)  p_\alpha p^\prime_\beta (m^2 - p' \cdot p).
 \label{eq.3.50}
\end{equation}
We can immediately conclude that $f_1 = f_2$ and $f_5 = f_6$. 
In order to reduce the number of terms further, we again contract with the 
polarization vectors and see that 
\begin{equation}
 \varepsilon^* (p';h') \cdot q =
 - \varepsilon^* (p';h') \cdot p, \quad
 \varepsilon (p;h) \cdot q =
 - \varepsilon (p;h) \cdot p'.
 \label{eq.3.60}
\end{equation}
So we are left with the term $(f_4 - f_3) p_\alpha p'_\beta$.
This structure is independent of the one containing $(f_5 - f_6)$,
because the latter originates from a term that contains the factor
$p^{\prime\mu} + p^\mu$ while the former does not. So we conclude that
$f_3 = f_4$, which means that only three independent form factors remain.

Next we impose hermiticity, {\it i.e.}
\begin{equation}
 \langle p';h'| J^\mu |p;h\rangle = \langle p;h| J^\mu |p';h'\rangle^* ,
 \label{eq.3.70}
\end{equation}
which gives after some rearrangement
\begin{equation}
 \varepsilon^{*\alpha} (p';h') J^\mu_{\alpha\beta} (p', p)
 \varepsilon^\beta (p;h) =
 \varepsilon^{*\alpha} (p';h') J^{\mu *}_{\beta\alpha} (p,p')
 \varepsilon^\beta (p;h) .
 \label{eq.3.80}
\end{equation}
This is an identity for all $p$, $p'$, $h$, and $h'$, so we find
\begin{equation}
 J^\mu_{\alpha\beta} (p', p) = J^{\mu *}_{\beta\alpha} (p,p').
 \label{eq.3.90}
\end{equation}
If we apply this identity to the structures we found, we see that the
coefficients of the tensors must be real, which means that $F_1$,
$F_2$, and $F_3$ in Eq.~(\ref{eq.3.10}) are real\footnote{Note that the
kinematic region for this discussion is spacelike, {\it i.e.} $q^2 <
0$.}.

The symmetry of $ J^\mu_{\alpha\beta} (p', p)$ entails relations between the matrix elements too. If we, in addition, apply Eq.~(\ref{eq.2.20}), which we
owe to the fact that the polarization vectors are standard spherical
tensors, we can deduce
\begin{equation}
 G^{\mu *}_{h'h} (p',p) = (-1)^{h' + h} G^\mu_{-h' -h} (p',p).
 \label{eq.3.100}
\end{equation}
The explicit expressions we are writing down in the next sections of course
satisfy these indentities.

Eq.~(\ref{eq.3.10}) can be split in an obvious way into the pieces
$J(1)F_1$, $J(2)F_2$, and $J(3)F_3$. Then we find for the polarization tensor
$G = G(1) F_1 + G(2) F_2 + G(3) F_3$ with the partial tensors
\begin{eqnarray}
 G^\mu_{h^\prime h}(1) & = & - (p^\prime + p)^\mu \;
 \varepsilon^*(p^\prime; h^\prime) \cdot \varepsilon(p; h),
 \nonumber \\
 G^\mu_{h^\prime h}(2) & = &
 - p^\prime \cdot \varepsilon(p; h)\; \varepsilon^{*\,\mu}(p^\prime; h^\prime)\;
 - p \cdot \varepsilon^*(p^\prime; h^\prime) \; \varepsilon^\mu(p; h) ,
 \nonumber \\
 G^\mu_{h^\prime h}(3) & = & - \frac{(p^\prime + p)^\mu}{2 m^2} \;
    p^\prime \cdot \varepsilon(p; h) \;
   p \cdot \varepsilon^*(p^\prime; h^\prime) ,
 \label{eq.3.110}
\end{eqnarray}
Clearly, we need three simple scalar products which we shall write in the
front form only
\begin{eqnarray}
 \varepsilon^*(p^\prime; 0) \cdot \varepsilon(p; 0) & = &
 \frac{p^{\prime \, +\, 2} (\vec{p}^{\,2}_\perp - m^2)
 + p^{+\, 2} (\vec{p}^{\,\prime \,2}_\perp - m^2)
 -2p^{\prime +} p^+ \vec{p}^{\,\prime}_\perp \cdot \vec{p}_\perp }
 {2m^2 p^{\prime +} p^+},
 \nonumber \\
 \varepsilon^*(p^\prime; 0) \cdot \varepsilon(p; h) & = &
 \frac{p^{\prime \, +} p^h - p^+ p^{\prime \, h}}{m p^+},
 \nonumber \\
 \varepsilon^*(p^\prime; h^\prime) \cdot \varepsilon(p; h) & = &
 -\frac{1+h^\prime h}{2} ,
 \nonumber \\
 \varepsilon^*(p^\prime; 0) \cdot p & = & 
 \frac{p^{\prime +\, 2} \vec{p}^{\,2}_\perp +
 p^{+\, 2} \vec{p}^{\,\prime 2}_\perp + m^2 (p^{\prime +\,2} - p^{+\,2})
 -2 p^{\prime +} p^+ \vec{p}^{\, \prime}_\perp  \cdot \vec{p}_\perp }
 {2m p^{\prime +} p^+},
 \nonumber \\
 \varepsilon^*(p^\prime; h) \cdot p & = &
 \frac{p^{\prime \, +} p^{-h} - p^+ p^{\prime \, -h}}{p^{\prime \, +}}.
 \label{eq.3.120}
\end{eqnarray}
where we made the obvious identification $p^{h=+1} \leftrightarrow p^r$,
$p^{h=-1} \leftrightarrow p^l$.

We give the matrix elements of the polarization tensors. We define 
$\tilde{G}^+(1)$ as $G^+(1) = (p^{\prime +}  + p^+)\, \tilde{G}^+(1)$ and find
\begin{eqnarray}
 \makebox[15mm][l]{$\tilde{G}^+(1)_{++}$} & = & \tilde{G}^+(1)_{--} = 1, \quad
 \makebox[15mm][l]{$\tilde{G}^+(1)_{+-}$}   =   \tilde{G}^+(1)_{-+} = 0, \nonumber \\
 \makebox[15mm][l]{$\tilde{G}^+(1)_{+0}$} & = & 
 \frac{p^+ p^{\prime l} - p^{\prime +} p^l}{mp^{\prime +}} , \nonumber \\
 \makebox[15mm][l]{$\tilde{G}^+(1)_{0+}$}& = &
 \frac{p^+ p^{\prime r} - p^{\prime +} p^r}{mp^+} , \nonumber \\
 \makebox[15mm][l]{$\tilde{G}^+(1)_{00}$}& = &
 -\frac{p^{\prime +\, 2} (\vec{p}^{\, 2}_\perp - m^2)
 + p^{+\, 2} (\vec{p}^{\, \prime 2}_\perp - m^2) -
 2p^{\prime +} p^+ \vec{p}^{\, \prime}_\perp \cdot \vec{p}_\perp}
 {2m^2 p^{\prime +} p^+} ,
 \nonumber \\
 \makebox[15mm][l]{$\tilde{G}^+(1)_{0-}$}& = &
 \frac{p^+ p^{\prime l} - p^{\prime +} p^l}{mp^+} , \nonumber \\
 \makebox[15mm][l]{$\tilde{G}^+(1)_{-0}$}& = &
 \frac{p^+ p^{\prime r} - p^{\prime +} p^r}{mp^{\prime +}}.
 \label{eq.3.130}
\end{eqnarray}
\begin{eqnarray}
 \makebox[15mm][l]{$G^+(2)_{++}$} & = & G^+(2)_{+-} = G^+(2)_{-+} =
 G^+(2)_{--} = 0, \nonumber \\
 \makebox[15mm][l]{$G^+(2)_{+0}$} & = &
 \frac{p^+(p^+ p^{\prime l} - p^{\prime +} p^l)} {mp^{\prime +}} ,
 \nonumber \\
 \makebox[15mm][l]{$G^+(2)_{0+}$} & = &\frac{p^{\prime +}
 (p^+ p^{\prime r} - p^{\prime +} p^r)}{mp^+} ,
 \nonumber \\
 \makebox[15mm][l]{$G^+(2)_{00}$} & = &
 \frac{p^{\prime +} +p^+}{2m^2 p^{\prime +} p^+} 
 [m^2 (p^{\prime +} - p^+)^2 - p^{\prime + \, 2} \vec{p}^{\, 2}_\perp
 -p^{+ \, 2} \vec{p}^{\, \prime 2}_\perp +
 2p^{\prime +} p^+ \vec{p}^{\, \prime}_\perp \cdot \vec{p}_\perp]  ,
 \nonumber \\
 \makebox[15mm][l]{$G^+(2)_{0-}$} & = &
 \frac{p^{\prime +} (p^+ p^{\prime l} - p^{\prime +} p^l)}{mp^+}  ,
 \nonumber \\
 \makebox[15mm][l]{$G^+(2)_{-0}$} & = &
 \frac{p^+ (p^+ p^{\prime r} - p^{\prime +} p^r)}{mp^{\prime +}} .
 \label{eq.3.140}
\end{eqnarray}
$G^+(3)$ also contains  an over-all factor, so we define
$G^+(3) = (p^{\prime +} + p^+)/(4 m^2 p^{\prime +} p^+) \tilde{G}^+(3)$ with
\begin{eqnarray}
 \makebox[15mm][l]{$\tilde{G}^+(3)_{++}$} & = &
 \makebox[15mm][l]{$\tilde{G}^+(3)_{--}$} =
 p^{\prime + \, 2} \vec{p}^{\, 2}_\perp + p^{+\, 2} \vec{p}^{\,\prime 2}_\perp
 - 2 p^{\prime +} p^+ \vec{p}^{\, \prime}_\perp \cdot \vec{p}_\perp ,
 \nonumber \\
 \makebox[15mm][l]{$\tilde{G}^+(3)_{+0}$} & = &
 \frac{p^+ p^{\prime l} - p^{\prime +} p^l}{m  p^{\prime +}} \;
 [p^{\prime + \, 2} \vec{p}^{\, 2}_\perp 
 +p^{+ \, 2} \vec{p}^{\,\prime 2}_\perp + m^2 (p^{+ \, 2} - p^{\prime + \, 2})
 -2 p^{\prime +} p^+ \vec{p}^{\, \prime}_\perp \cdot \vec{p}_\perp ] ,
 \nonumber \\
 \makebox[15mm][l]{$\tilde{G}^+(3)_{+-}$} & = &
 -2 (p^+ p^{\prime l} - p^{\prime +} p^l)^2  ,
 \nonumber \\
 \makebox[15mm][l]{$\tilde{G}^+(3)_{0+}$} & = &
 \frac{p^+ p^{\prime r} - p^{\prime +} p^r}{m  p^+} \;
 [p^{\prime + \, 2} \vec{p}^{\, 2}_\perp 
 +p^{+\, 2} \vec{p}^{\,\prime 2}_\perp - m^2 (p^{+\, 2} - p^{\prime + \, 2})
 -2 p^{\prime +} p^+ \vec{p}^{\, \prime}_\perp  \cdot \vec{p}_\perp] ,
 \nonumber \\
 \makebox[15mm][l]{$\tilde{G}^+(3)_{00}$} & = &
 \frac{-1}{2m^2 p^+ p^{\prime +}} \;
 [\,(p^{\prime +\, 2} \vec{p}^{\,2}_\perp + p^{+\, 2} \vec{p}^{\,\prime 2}_\perp
 - m^2 (p^{+\,2} - p^{\prime +\,2})
 -2 p^{\prime +} p^+ \vec{p}^{\, \prime}_\perp  \cdot \vec{p}_\perp )
 \nonumber \\
 & & \quad \quad \quad \quad \; \times
 (p^{\prime +\,2} \vec{p}^{\,2}_\perp + p^{+\,2} \vec{p}^{\,\prime 2}_\perp
 + m^2 (p^{+\,2} - p^{\prime +\,2})
 -2 p^{\prime +} p^+ \vec{p}^{\, \prime}_\perp  \cdot \vec{p}_\perp )\, ] ,
 \nonumber \\
 \makebox[15mm][l]{$\tilde{G}^+(3)_{0-}$} & = &
 \frac{p^+ p^{\prime l} - p^{\prime +} p^l}{m  p^+} \;
 [p^{\prime +\,2} \vec{p}^{\,2}_\perp 
 + p^{+\,2} \vec{p}^{\,\prime 2}_\perp - m^2 (p^{+\,2} - p^{\prime +\,2})
 -2 p^{\prime +} p^+ \vec{p}^{\, \prime}_\perp  \cdot \vec{p}_\perp ] ,
 \nonumber \\
 \makebox[15mm][l]{$\tilde{G}^+(3)_{-+}$} & = &
 -2 (p^{\prime +} p^r - p^+ p^{\prime r})^2  ,
 \nonumber \\
 \makebox[15mm][l]{$\tilde{G}^+(3)_{-0}$} & = &
 \frac{p^+ p^{\prime r} - p^{\prime +} p^r}{m  p^{\prime +}} \;
 [p^{\prime +\,2} \vec{p}^{\,2}_\perp 
 + p^{+\,2} \vec{p}^{\,\prime 2}_\perp + m^2 (p^{+\,2} - p^{\prime +\,2})
 -2 p^{\prime +} p^+ \vec{p}^{\, \prime}_\perp  \cdot \vec{p}_\perp )\, ] . 
 \nonumber \\
 \label{eq.3.150}
\end{eqnarray}
Hermiticity follows from the simultaneous replacements 
$p \leftrightarrow p^\prime$ and $p^l \leftrightarrow -p^r$. 

\subsection{Symmetries of the Polarization Tensor}
\label{sec.III.A}
The formulae above tell us that the polarization tensor has the following form
\begin{equation}
 G(i) = \left(
 \begin{array}{ccc}
 a_i & c_i & e^*_i \\ b_i & d_i & -b^*_i \\ e_i & -c^*_i & a_i
 \end{array}
 \right),
 \label{eq.3.160}
\end{equation}
which is valid for all three contributions $G(i)$, $i=1,2,3$. Using an
obvious notation we find for the complete polarization tensor the form
\begin{eqnarray}
 G & = & \left(
 \begin{array}{ccc}
 a_1 F_1 + a_3 F_3 & c_1 F_1 + c_2 F_2 + c_3 F_3 & e^*_3 F_3 \\
 b_1 F_1 + b_2 F_2 + b_3 F_3 & d_1 F_1 + d_2 F_2 + d_3 F_3 &
 -(b_1 F_1 + b_2 F_2 + b_3 F_3)^* \\
 e_3 F_3 & -(c_1 F_1 + c_2 F_2 + c_3 F_3)^* & a_1 F_1 + a_3 F_3 \end{array}
 \right)
 \nonumber \\
  & = & \left(
 \begin{array}{ccc}
 G^+_{++} & G^+_{+0} & G^+_{+-} \\
 G^+_{0+} & G^+_{00} & G^+_{0-} \\
 G^+_{-+} & G^+_{-0} & G^+_{--}
 \end{array}
 \right).
 \label{eq.3.170}
\end{eqnarray}

Apparently, the tensor components we obtain here satisfy an additional
identity
\begin{equation}
 G^+_{++} = G^+_{--} =  G^{+*}_{++}.
 \label{eq.3.180}
\end{equation}
This result is specific for the choice of $\mu$: it is true for the
good current $J^+$, but does not apply to the terrible current $J^-$.
The matrix elements $G^-_{++}$ and $G^-_{--}$ are not real, but they are
complex conjugates.

For later reference we also give the expressions for $a_1, \dots, e_3$.
\begin{eqnarray}
 a_1 & = & \quad p^{\prime\, +} + p^+ ,
 \nonumber \\
 b_1 & = & - (p^{\prime\, +} + p^+) \,
 \frac{p^{\prime\, +} p^r - p^+ p^{\prime\, r}}{m p^+},
 \nonumber \\
 c_1 & = & - (p^{\prime\, +} + p^+) \,
 \frac{p^{\prime\, +} p^l - p^+ p^{\prime\, l}}{m p^{\prime\, +}},
 \nonumber \\
 d_1 & = & \quad (p^{\prime\, +} + p^+) \,
 \frac{m^2 (p^{\prime\, +\, 2} + p^{+\, 2}) +
 2 (p^{\prime\, +} p^r - p^+ p^{\prime\, r})
 (p^{\prime\, +} p^l - p^+ p^{\prime\, l})}
 {2 m^2 p^{\prime\, +} p^+} ,
 \nonumber \\
 e_1 & = & 0.
 \label{eq.3.190}
\end{eqnarray}
\begin{eqnarray}
 a_2 & = & \quad 0,
 \nonumber \\
 b_2 & = & - p^{\prime\, +} \,
 \frac{p^{\prime\, +} p^r - p^+ p^{\prime\, r}}{mp^+} ,
 \nonumber \\
 c_2 & = & - p^+ \,
 \frac{p^{\prime\, +} p^l - p^+ p^{\prime\, l}}{mp^{\prime\, +}} ,
 \nonumber \\
 d_2 & = & \quad (p^{\prime\, +} + p^+) \,
 \frac{m^2 (p^{\prime\, +} - p^{+})^2 +
 2 (p^{\prime\, +} p^r - p^+ p^{\prime\, r})
 (p^{\prime\, +} p^l - p^+ p^{\prime\, l})}
 {2 m^2 p^{\prime\, +} p^+} ,
 \nonumber \\
 e_2 & = & \quad 0.
 \label{eq.3.200}
\end{eqnarray}
\begin{eqnarray}
 a_3 & = &  - (p^{\prime\, +} + p^+) \,
 \frac{(p^{\prime\, +} p^r - p^+ p^{\prime\, r})
 (p^{\prime\, +} p^l - p^+ p^{\prime\, l})}{2m^2 p^{\prime\, +} p^+} ,
 \nonumber \\
 b_3 & = & - (p^{\prime\, +} + p^+) \,
 \frac{(p^{\prime\, +} p^r - p^+ p^{\prime\, r})
 [m^2 (p^{\prime\, +\, 2} -p^{+\, 2}) -
 2 (p^{\prime\, +} p^r - p^+ p^{\prime\, r})
 (p^{\prime\, +} p^l - p^+ p^{\prime\, l})]}
 {4 m^3 p^{\prime\, +} p^{+\, 2} } ,
 \nonumber \\
 c_3 & = & \quad (p^{\prime\, +} + p^+) \,
 \frac{(p^{\prime\, +} p^l - p^+ p^{\prime\, l})
 [m^2 (p^{\prime\, +\, 2} -p^{+\, 2}) +
 2 (p^{\prime\, +} p^r - p^+ p^{\prime\, r})
 (p^{\prime\, +} p^l - p^+ p^{\prime\, l})]}
 {4 m^3 p^{\prime\, +\, 2} p^{+} } ,
 \nonumber \\
 d_3 & = & \quad \frac{p^{\prime\, +} + p^+}{8m^4 p^{\prime\, +\, 2} p^{+\, 2}}
 \; [m^2 (p^{\prime\, +\, 2} -p^{+\, 2}) -
 2 (p^{\prime\, +} p^r - p^+ p^{\prime\, r})
 (p^{\prime\, +} p^l - p^+ p^{\prime\, l})]
 \nonumber \\
 && \quad \quad \quad \quad \quad \quad
 \times [m^2 (p^{\prime\, +\, 2} -p^{+\, 2}) +
 2 (p^{\prime\, +} p^r - p^+ p^{\prime\, r})
 (p^{\prime\, +} p^l - p^+ p^{\prime\, l})] ,
 \nonumber \\
 e_3 & = & - (p^{\prime\, +} + p^+) \,
 \frac{(p^{\prime\, +} p^r - p^+ p^{\prime\, r})^2}
 {2m^2 p^{\prime\, +} p^+} .
 \label{eq.3.210}
\end{eqnarray}

We see that the nine matrix elements of $G$ have four relations that 
involve a phase factor only, viz
\begin{equation}
 G^+_{++} = G^+_{--}, \quad G^+_{+-} =G^{+*}_{-+}, \quad G^+_{0-} 
=-G^{+*}_{0+}, \quad
 G^+_{+0} = -G^{*+}_{-0}.
 \label{eq.3.220}
\end{equation}
We need two more equations that express the fact that there are only
three independent form factors. These consistency conditions are the two
angular conditions proper. Since we are working only with the + component
of the current, we shall use the following short-hand notations
\[
 G_a = G^+_{++} = G^{+*}_{--}, \quad G_b = G^+_{0+} = -G^{+*}_{0-}, 
\]
\begin{equation}
 G_c = G^+_{+0} = -G^{+*}_{-0}, \; G_d = G^+_{00}, \;
 G_e = G^+_{-+} = G^{+*}_{+-}.
 \label{eq.3.230}
\end{equation}

We can now solve for $F_i$ in an obvious way. First we obtain $F_3$ from 
$G_e$, then $F_1$ from $G_a$ and $F_3$. Then we have a choice whether we 
want to obtain $F_2$ from $G_b$, $G_c$ or $G_d$; these solutions we denote by
$F^b_2$, $F^c_2$, and $F^d_2$, respectively. 
As these results must coincide, the identity of these three results form the
angular conditions: $F^b_2 = F^c_2 = F^d_2$. We find

\begin{eqnarray}\
 F_1 & = & \frac{1}{a_1} G_a - \frac{a_3}{a_1 e_3} G_e , \nonumber \\
 F_3 & = & \frac{1}{e_3} G_e, \nonumber \\
 F^b_2 & = & \frac{1}{b_2}
 \left[ -\frac{b_1}{a_1} \, G_a + G_b +
 \frac{a_3 b_1 - a_1 b_3}{a_1 e_3} \, G_e \right], \nonumber \\
 F^c_2 & = & \frac{1}{c_2}
 \left[ -\frac{c_1}{a_1} \, G_a + G_c +
 \frac{a_3 c_1 - a_1 c_3}{a_1 e_3} \, G_e \right], \nonumber \\
 F^d_2 & = & \frac{1}{d_2}
 \left[ -\frac{d_1}{a_1} \, G_a + G_d +
 \frac{a_3 d_1 - a_1 d_3}{a_1 e_3} \, G_e \right].
 \label{eq.3.240}
\end{eqnarray}

The relations Eq.~(\ref{eq.3.100}) reduce the nine complex elements
of the polarization tensor to nine real numbers. As there are only three
real independent form factors, we need six linear relations to realize the reduction from nine to three. The equations above serve this purpose. By
equating the real and imaginary parts of the two sides of the first three
of Eqs.~(\ref{eq.3.240}), we find six relations that must hold for
the components of $G^\mu_{h'h}$. Having thus achieved the reduction to the
minimum number of independent functions, the other equations must be
considered to be {\em consistency conditions}. As the three equations
expressing $F_2$ in terms of the tensor components are not independent, but
form a system of rank two, only one complex equation, or two real ones
remain.

In the literature usually only one is given, said to be the {\em angular
condition}. From our considerations it must be clear that there are indeed
two conditions.

\subsection{Angular Conditions}
\label{sec.III.B}
The angular conditions can now be formulated succinctly:
\begin{equation}
 F^b_2 = F^c_2, \quad F^b_2 = F^d_2, \quad F^c_2 = F^d_2.
 \label{eq.3.250}
\end{equation}
We shall write these conditions explicitly for unspecified kinematics. 

The first one, denoted henceforth by AC 1, is
\begin{eqnarray}
 F^b_2 - F^c_2 & = &  0 \nonumber \\
 & = & \frac{p^{\prime \, +} - p^+}{p^{\prime \, +} p^+}
 \left[ G_a + \frac{m^2}{2} \frac{(p^{\prime \, +} + p^+)^2}
 {(p^{\prime \, +} p^r - p^+ p^{\prime\, r})^2} G_e \right]
 \nonumber \\
 &&
 - m \, \frac{p^+}{p^{\prime \, +}}
 \frac{1}{p^{\prime \, +} p^r - p^+ p^{\prime\, r}} \; G_b
 + m \, \frac{p^{\prime \, +}}{p^+}
 \frac{1}{p^{\prime \, +} p^l - p^+ p^{\prime\, l}} \; G_c .
 \label{eq.3.260}
\end{eqnarray}
The second one, AC 2, is
\begin{eqnarray}
 F^b_2 - F^d_2 & = &  0 \nonumber \\
 & = & \left[-\frac{1}{p^{\prime \, +}} +
 \frac{m^2 (p^{\prime \, +\, 2} + p^{+\, 2})
 + 2(p^{\prime \, +} p^r - p^+ p^{\prime\, r})
 (p^{\prime \, +} p^l - p^+ p^{\prime\, l}) }
 {(p^{\prime \, +} + p^+)[m^2(p^{\prime \, +} - p^+)^2
 + 2(p^{\prime \, +} p^r - p^+ p^{\prime\, r})
 (p^{\prime \, +} p^l - p^+ p^{\prime\, l})]} \right] G_a
 \nonumber \\
 &&
 - m \frac{p^+}{p^{\prime \, +}}
 \frac{1}{p^{\prime \, +} p^r - p^+ p^{\prime\, r}} \; G_b
 \nonumber \\
 &&
 - \frac{2m^2 p^{\prime \, +} p^+}{p^{\prime \, +} + p^+} \,
 \frac{1}{m^2(p^{\prime \, +} - p^+)^2
 + 2(p^{\prime \, +} p^r - p^+ p^{\prime\, r})
 (p^{\prime \, +} p^l - p^+ p^{\prime\, l})} \; G_d
 \nonumber \\
 &&
 + \left[\frac{m^2}{2p^{\prime \, +}} \frac{p^{\prime \, +\, 2} - p^{+ \, 2}}
 {(p^{\prime \, +} p^r - p^+ p^{\prime\, r})^2} 
 - \frac{m^2}
 {2(p^{\prime \, +} + p^+)(p^{\prime \, +} p^r - p^+ p^{\prime\, r})^2}
 \right.
 \nonumber \\
 && \left. \times
 \frac{m^2 (p^{\prime \, +\,2} - p^{+\,2})^2 + 2(p^{\prime\, +\, 2} + p^{+\,2}) 
 (p^{\prime \, +} p^r - p^+ p^{\prime\, r})
 (p^{\prime \, +} p^l - p^+ p^{\prime\, l}) }
 {m^2(p^{\prime \, +} - p^+)^2
 + 2(p^{\prime \, +} p^r - p^+ p^{\prime\, r})
 (p^{\prime \, +} p^l - p^+ p^{\prime\, l})]} \right] \; G_e. \nonumber \\
 \label{eq.3.270}
\end{eqnarray}
The last one is
\begin{eqnarray}
 F^c_2 - F^d_2 & = &  0 \nonumber \\
 & = & \left[-\frac{1}{p^+} +
 \frac{m^2 (p^{\prime \, +\, 2} + p^{+\, 2})
 + 2(p^{\prime \, +} p^l - p^+ p^{\prime\, l})
 (p^{\prime \, +} p^l - p^+ p^{\prime\, l}) }
 {(p^{\prime \, +} + p^+)[m^2(p^{\prime \, +} - p^+)^2
 + 2(p^{\prime \, +} p^r - p^+ p^{\prime\, r})
 (p^{\prime \, +} p^l - p^+ p^{\prime\, l})]} \right] \; G_a
 \nonumber \\
 &&
 - m \frac{p^{\prime \, +}}{p^+}
 \frac{1}{p^{\prime \, +} p^r - p^+ p^{\prime\, r}} \; G_c
 \nonumber \\
 &&
 - \frac{2m^2 p^{\prime \, +} p^+}{p^{\prime \, +} + p^+} \,
 \frac{1}{m^2(p^{\prime \, +} - p^+)^2
 + 2(p^{\prime \, +} p^r - p^+ p^{\prime\, r})
 (p^{\prime \, +} p^l - p^+ p^{\prime\, l})} \; G_d
 \nonumber \\
 &&
 +  \left[- \frac{m^2}{2p^+} \frac{p^{\prime \, +\, 2} - p^{+ \, 2}}
 {(p^{\prime \, +} p^r - p^+ p^{\prime\, r})^2}
 - \frac{m^2}
 {2(p^{\prime\, +} + p^+)(p^{\prime \, +} p^r - p^+ p^{\prime\, r})^2}
 \right.
 \nonumber \\
 && \left. \times
 \frac{m^2 (p^{\prime\, +\, 2} -p^{+\, 2})^2 + 2(p^{\prime\, +\, 2} + p^{+\, 2})
 (p^{\prime \, +} p^r - p^+ p^{\prime\, r})
 (p^{\prime \, +} p^l - p^+ p^{\prime\, l}) }
 {m^2(p^{\prime \, +} - p^+)^2
 + 2(p^{\prime \, +} p^r - p^+ p^{\prime\, r})
 (p^{\prime \, +} p^l - p^+ p^{\prime\, l})]} \right] \; G_e. \nonumber \\
 \label{eq.3.280}
\end{eqnarray}

If we substitute Eq.~(\ref{eq.3.260}) into Eq.~(\ref{eq.3.270}) we see that
it is equivalent to Eq.~(\ref{eq.3.280}), as it must be, because these
equations are not independent as there are only two independent angular
conditions.

Clearly, these conditions are quite complicated. We can simplify them by
factoring out some common factors, at the same time avoiding denominators that
may vanish. Instead of Eqs.~(\ref{eq.3.260},\ref{eq.3.270}) we get the 
conditions AC 1
\begin{eqnarray}
 2(p^{\prime \, +} - p^+)
 (p^{\prime \, +} p^r - p^+ p^{\prime\, r})^2 
 (p^{\prime \, +} p^l - p^+ p^{\prime\, l}) \; G_a 
 &&
 \nonumber \\
 -2m p^{+\, 2}(p^{\prime \, +} p^r - p^+ p^{\prime\, r})
 (p^{\prime \, +} p^l - p^+ p^{\prime\, l}) \; G_b
 &&
 \nonumber \\
 +2m p^{\prime \, +\, 2} (p^{\prime \, +} p^r - p^+ p^{\prime\, r})^2 \; G_c
 &&
 \nonumber \\
 + m^2 (p^{\prime \, +} - p^+) (p^{\prime \, +} + p^+)^2
 (p^{\prime \, +} p^l - p^+ p^{\prime\, l}) \; G_e & = & 0.
 \label{eq.3.290}
\end{eqnarray}
and AC 2
\begin{eqnarray}
 2 (p^{\prime \, +} p^r - p^+ p^{\prime\, r})^2 
 [m^2 (p^{+\, 2} -2 p^{\prime \, +} p^+ - p^{\prime \, +\, 2})
 +2 (p^{\prime \, +} p^r - p^+ p^{\prime\, r}) 
 (p^{\prime \, +} p^l - p^+ p^{\prime\, l}) ] \; G_a
 &&
 \nonumber \\
 +2m (p^{\prime \, +} + p^+)(p^{\prime \, +} p^r - p^+ p^{\prime\, r})
 [m^2 (p^{\prime \, +} - p^+)^2
 +2 (p^{\prime \, +} p^r - p^+ p^{\prime\, r}) 
 (p^{\prime \, +} p^l - p^+ p^{\prime\, l}) ]  \; G_b
 &&
 \nonumber \\
 +4m^2 p^{\prime \, +\, 2} (p^{\prime \, +} p^r - p^+ p^{\prime\, r})^2 \; G_d
 &&
 \nonumber \\
 + m^2 [m^2 (p^{\prime \, +\, 2} - p^{+\, 2})^2
 +2 (p^{+\, 2} +2 p^{\prime \, +} p^+ - p^{\prime \, +\, 2})
 (p^{\prime \, +} p^r - p^+ p^{\prime\, r})
 (p^{\prime \, +} p^l - p^+ p^{\prime\, l})] \; G_e & = & 0.
 \nonumber \\
 \label{eq.3.300}
\end{eqnarray}

Clearly, these conditions are minimal, as we cannot eliminate any of the
five tensor components to obtain a simpler one. 

It is useful to realize the phase relations that occur. Besides the
relations expressed in Eq.~(\ref{eq.3.160}, \ref{eq.3.220}) we can use the
fact that $(p^l)^* = -p^r$ and the fact that $G_a$ and $G_d$ are real
to infer that both angular conditions have the form 
\begin{equation}
 C_a \, G_a + C_b e^{-i \phi} \, G_b + C_c e^{i \phi} \, G_c + C_d \,
 G_d + C_e e^{-2i \phi} G_e = 0,
 \label{eq.3.310} 
\end{equation} 
where the coefficients $C_a, \dots C_e$ are real and 
$\phi$ is the argument of the complex number $p^{\prime \, +} p^r - p^+
p^{\prime\, r}$, given by
\begin{equation}
 \tan \phi = - \frac{p^+ p'_y - p^{\prime +} p_y}{p^+ p'_x - p^{\prime +} 
p_x}.
 \label{eq.3.320}
\end{equation}
This angle can be set to zero by a rotation of the
reference frame about the $z$-axis.
This rotation being kinematical in LFD we may expect the phase relations to 
be satisfied always.

It may turn out for some kinematics, that these relations simplify.
This happens to be the case in {\it e.g.} the DYW-frame, where $p^{\prime \,
+} = p^+$ and $\vec{p}_\perp = 0$. Moreover, when $\vec{q}$ is purely 
longitudinal, {\it i.e.}, $\vec{q}_\perp = 0$, we can rotate the reference frame
such that $\vec{p}_\perp = \vec{p}^{\,\prime}_\perp = 0$. Then both
angular conditions are identically satisfied, as all coefficients vanish.

\section{Specific Frames}
\label{sec.IV}
We consider three specific frames: the Drell-Yan-West (DYW), Breit and
Target-Rest (TRF) frames. For simplicity, only the kinematics 
and the angular conditions in the form $F^b_2 - F^c_2 = 0$ (AC 1 )and
$F^b_2 -F^d_2 =0$ (AC 2) are presented in this section and
the detailed formulas of the polarization tensors in the form 
of the coefficients $a_1, \dots, e_3$ are summarized in Appendix~\ref{sec.C}.

\subsection{Drell-Yan-West Frame}
\label{sec.IV.A}
\subsubsection{Kinematics}
For the DYW frame,
\begin{eqnarray}
 p & = & (p^+, 0,0, m^2/(2 p^+)) \nonumber \\
 q & = & (0, q_x, q_y, \vec{q}^{\,2}_\perp/(2 p^+)) \nonumber \\
 p^\prime & = & p + q \nonumber \\
 & = & (p^+, q_x, q_y, (\vec{q}^{\,2}_\perp+m^2)/(2 p^+)),
 \label{eq.4.10}
\end{eqnarray}
with the identification $q_x = Q \cos \phi$,
$q_y = Q \sin \phi$ one finds the explicit formulas
\begin{eqnarray}
 p & = & (p^+, 0,0, m^2/(2 p^+)) \nonumber \\
 q & = & (0, Q  \cos \phi,Q  \sin \phi, Q^2 /(2 p^+))
 \nonumber \\
 p^\prime 
 & = & (p^+, Q \cos \phi, Q \sin \phi, (Q^2 + m^2)/(2 p^+)) 
 \label{eq.4.20}
\end{eqnarray}
and
\begin{equation}
 q^r = -\frac{Q}{\surd 2} e^{i \phi}, \quad
 q^l = \frac{Q}{\surd 2} e^{-i \phi}.
 \label{eq.4.30}
\end{equation}

\subsubsection{Angular Conditions}
We write the angular conditions mentioned in the  previous section. 

\noindent AC 1
\begin{equation}
  e^{-i \phi} G_b + e^{i \phi} G_c  = 0,
 \label{eq.4.40}
\end{equation}
AC 2
\begin{equation}
 (2m^2 + Q^2) \, G_a
 + 2\surd 2 m Q e^{-i \phi} \, G_b
 - 2 m^2  \, G_d + 2 m^2 e^{-2i \phi} \, G_e = 0.
 \label{eq.4.50}
\end{equation}

\subsection{Breit Frame}
\label{sec.IV.B}
\subsubsection{Kinematics}

We define the quantity $\beta$ as
\begin{equation}
 \beta = \sqrt{1+\left(\frac{Q}{2m}\right)^2}.
 \label{eq.4.60}
\end{equation}
Then
\begin{eqnarray}
 p & = &
 \left(\frac{2 m \beta - Q \cos \theta}{2 \surd 2},
         -\frac{Q \sin \theta \cos \phi}{2},
         -\frac{Q \sin \theta \sin \phi}{2},
       \frac{2 m \beta + Q\cos \theta}{2\surd 2}
 \right),
 \nonumber \\
 p^\prime & = &
 \left(\frac{2 m \beta + Q \cos \theta}{2 \surd 2},
         \frac{Q \sin \theta \cos \phi}{2},
         \frac{Q \sin \theta \sin \phi}{2},
       \frac{2 m \beta - Q\cos \theta}{2\surd 2}
 \right),
 \nonumber \\
 q & = &
 \left(\frac{Q \cos \theta}{\surd 2},
       Q \sin \theta \cos \phi,
       Q \sin \theta \sin \phi,
       -\frac{Q \cos \theta}{\surd 2} \right). 
 \label{eq.4.70}
\end{eqnarray}

\subsubsection{Angular conditions}
\label{sec.IV.B.3}
By now we give only the two linearly independent conditions.
We simplify the expressions as much as possible by dividing out common
factors to find the two conditions. 

\noindent AC 1
\begin{eqnarray}
 - 2\surd 2 \beta Q^2 \cos\theta \sin^2 \theta \; G_a &&
 \nonumber \\
 + (2 \beta m - Q \cos\theta )^2 \sin\theta e^{-i \phi} \; G_b &&
 \nonumber \\
 + (2 \beta m + Q \cos\theta )^2 \sin\theta e^{i \phi} \; G_c &&
 \nonumber \\
 -8 \surd 2 \beta m^2 \cos\theta e^{-2i \phi} \; G_e & = & 0,
 \label{eq.4.80}
\end{eqnarray}
AC 2
\begin{eqnarray}
 - [4 \beta m Q \cos\theta - Q^2 \cos^2 \theta 
 + 2 \beta^2 (2 m^2 + Q^2 \sin^2 \theta)] \sin^2 \theta \; G_a &&
 \nonumber \\
 - 4\surd 2 m Q (\beta^2 \sin^2 \theta - \cos^2 \theta) \sin\theta
 e^{-i \phi} \; G_b &&
 \nonumber \\
 + (2 \beta m + Q \cos\theta )^2 \sin^2 \theta \; G_d &&
 \nonumber \\
 + [ (8m^2 + Q^2 \sin^2 \theta) \cos^2 \theta 
 +4 \beta m Q \cos\theta \sin^2 \theta
 - 4 \beta^2 m^2 \sin^2 \theta ] e^{-2i \phi} \; G_e & = & 0.
 \label{eq.4.90}
\end{eqnarray}
We note that Eqs.~(\ref{eq.4.80}) and (\ref{eq.4.90}) are reduced
to the results in DYW Eqs.~(\ref{eq.4.40}) and (\ref{eq.4.50}), respectively,
if $\theta = \pi/2$ as they should, because the two frames are
related by a kinematical transformation in that case and the angular 
conditions do not change under any kinematical transformation.

\subsection{Target-Rest Frame}
\label{sec.IV.C}
\subsubsection{Kinematics}
Using again $\beta$, and $\kappa$, defined as
\begin{equation}
 \kappa = \frac{Q^2}{2 m},
 \label{eq.4.100}
\end{equation}
we find
\begin{eqnarray}
 p & = & \left(\frac{m}{\surd 2}, 0,0, \frac{m}{\surd 2} \right). \nonumber \\
 q & = &
 \left( \frac{\kappa + \beta Q \cos \theta}{\surd 2}, 
 \beta Q \sin \theta \cos \phi,
 \beta Q \sin \theta \sin \phi,
 \frac{\kappa - \beta Q \cos \theta}{\surd 2}
 \right) , \nonumber \\
 p^\prime & = & p + q \nonumber \\
   & = & 
 \left(\frac{m + \kappa + \beta Q \cos \theta}{\surd 2},
 \beta Q \sin \theta \cos \phi,
 \beta Q \sin \theta \sin \phi,
 \frac{m + \kappa - \beta Q \cos \theta}{\surd 2} \right) .
 \label{eq.4.110}
\end{eqnarray}

\subsubsection{Angular conditions}
\label{sec.IV.C.4}
We give again only the two conditions after simplification by
dividing out as many common factors as possible.

\noindent AC 1
\begin{eqnarray}
 - \beta^2 Q^2 (\kappa + \beta Q \cos\theta) \sin^2 \theta \; G_a &&
 \nonumber \\
 + \surd 2 \beta m^2 Q \sin\theta \, e^{-i \phi} \; G_b &&
 \nonumber \\
 + \surd 2 \beta Q (m+\kappa + \beta Q \cos\theta )^2 \sin\theta \, e^{i \phi}
 \; G_c &&
 \nonumber \\
 -(\kappa + \beta Q \cos\theta )
 (2m + \kappa + \beta Q \cos\theta)^2 \, e^{-2i \phi} \; G_e & = & 0,
 \label{eq.4.120}
\end{eqnarray}
AC 2
\begin{eqnarray}
 - \beta^2 Q^2 [\kappa^2 + 4\kappa m + 2 m^2 + \beta^2 Q^2 
 + 2\beta (2m + \kappa) Q \cos\theta] \sin^2 \theta \; G_a &&
 \nonumber \\
 +\surd 2 \beta Q (2 m + \kappa + \beta Q \cos\theta)
 [\kappa^2 + 2 \beta \kappa Q \cos\theta 
 + \beta^2 Q^2 \cos 2 \theta]
 \sin\theta \, e^{-i \phi} \; G_b &&
 \nonumber \\
 + 2 \beta^2 Q^2 (m + \kappa + \beta Q \cos\theta)^2 
 \sin^2 \theta \; G_d &&
 \nonumber \\
 + [ (\kappa + \beta Q \cos\theta)^2
 (2m + \kappa + \beta Q \cos\theta)^2 \quad \quad \quad &&
 \nonumber \\
 + \beta^2 Q^2 (\kappa^2 - 2m^2 + 2 \beta\kappa Q \cos\theta
 + \beta^2 Q^2 \cos^2 \theta)] \sin^2 \theta \, e^{-2i \phi} \; G_e & = & 0.
 \nonumber \\
 \label{eq.4.130}
\end{eqnarray}
We note that Eqs.~(\ref{eq.4.120}) and (\ref{eq.4.130}) are identical
to Eqs.~(\ref{eq.4.40}) and (\ref{eq.4.50}) if $\beta \sin\theta =1$.

\section{Limiting Cases}
\label{sec.V}

In order to be able to interprete the angular conditions, we studied the
dependence on $Q$ in the limits $Q \to 0$ and $Q \to \infty$. We shall use
the notation
\begin{eqnarray}
 {\rm AC} \; 1 & \Longleftrightarrow &
 R^1_a \, G_a + R^1_b \, G_b + R^1_c \, G_c + R^1_e \, G_e = 0,
 \nonumber \\
 {\rm AC} \; 2 & \Longleftrightarrow &
 R^2_a \, G_a + R^2_b \, G_b + R^2_d \, G_d + R^2_e \, G_e = 0.
 \label{eq.5.10}
\end{eqnarray}

\subsection{ $Q \to 0$ Limit }

Using the definition of the physical form factors at $Q^2 =0$, {\it i.e.}
\begin{equation}
 eG_C(0) = e, \; eG_M(0) = 2m\mu_1 , \; eG_Q(0) = m^2 Q_1,
 \label{eq.5.20}
\end{equation}
we find from Eq.~(\ref{eq.3.20}) 
\begin{equation}
 F_1(0) = 1, \; F_2(0) = -\frac{2m\mu_1}{e}, 
 \; F_3(0) = -1 + \frac{2m\mu_1}{e} + \frac{m^2 Q_1}{e}.
 \label{eq.5.30}
\end{equation}
According to the universality condition given by Eq.~(\ref{eq.1.10}),
in the limit of bound-state radius $R \to 0$ the form factors
$F_i(0)$ for $i=2,3$ are reduced to 
\begin{equation}
 F_2(0)=-2, \; F_3(0)=0.
 \label{eq.5.40}
\end{equation}

Since the target is intact in the $Q \to 0$ limit,
$p^\mu = {p^\prime}^\mu$ and thus 
one can easily see from Eqs.~(\ref{eq.3.190})-(\ref{eq.3.210})
that $a_i = d_i$ and $b_i = c_i = e_i =0$ for all $i=1,2,3$.
Thus, we find $G_a = G_d$ or $G^+_{++} = G^+_{00}$ and all other
spin-flip amplitudes vanish in this limit regardless of reference frames.
This can be understood because the spin would not flip
if the target is intact and also the direction of spin 
wouldn't matter in this limit.
Moreover, all the coefficients ($R^i_a$, etc.) in Eq.~(\ref{eq.5.10})
vanish in $Q \to 0$ limit and thus both angular conditions, AC 1 
and AC 2, are trivially satisfied.

\subsection{Behavior for $Q \to \infty$}
\label{sec.V.A}
Imposing a naturalness condition, namely all three terms
in Eq.~(\ref{eq.3.10}) should be of the same order in $Q$, 
one can find that the form factors $F_i(Q^2)$ behave as
$F_1(Q^2) \sim F_2(Q^2) \sim \frac{Q^2}{m^2}F_3(Q^2)$
in the large $Q^2$ limit. Using this, we can derive
high $Q^2$ behaviors of the helicity amplitudes $G^+_{h'h}$
and the coefficients ($R^i_a$, etc.) of the angular conditions. 
In table \ref{tab.01}, we summarize the results.
\begin{table}
\caption{Leading behavior for $Q \to \infty$ the tensor components $G_a 
\dots G_e$ and the
coefficients $R^1_a \dots R^2_e$ in the different reference frames
considered. The BF and TRF are kinematically connected to the DYW 
frame only in particular angles $\theta_{BF} = \pi/2$ and 
$\theta_{TRF} = \theta_0 = \sin^{-1}(1/\beta)$, respectively.} 
\label{tab.01} %
\begin{tabular}{c|ccccccc}
 & \multicolumn{7}{c}{$Q \to \infty$} \\
 & DYW & \multicolumn{3}{c}{Breit}  & \multicolumn{3}{c}{TRF} \\ 
 &     & $\theta\neq 0,\pi/2$ & $\theta =\pi/2$ & 
$\theta=0$ & $\theta \neq 0,\theta_0$ & $\theta =0$ & 
$\theta = \theta_0$ \\ 
\hline
 \rule{1mm}{0mm} $G_a$ \rule{1mm}{0mm} 
       & 1 & $Q$ & $Q$ & $Q$ & $Q^2$ & $Q^2$ & $1$ \\
 $G_b$ & $Q$ & $Q^2$ & $Q^2$ & 0 & $Q^4$ & 0 & $Q$ \\
 $G_c$ & $Q$ & $Q^2$ & $Q^2$ & 0 & $Q^2$ & 0 & $Q$ \\
 $G_d$ & $Q^2$ & $Q^3$ & $Q^3$ & $Q^3$ & $Q^4$ & $Q^4$ & $Q^2$ \\
 $G_e$ & 1 & $Q$ & $Q$ & 0 & $Q^2$ & 0 & 1 \\
\hline
 $R^1_a$ & 0 & $Q^3$ & 0 & 0 & $Q^6$ & 0 & 0 \\
 $R^1_b$ & 1 & $Q^2$ & $Q^2$ & 0 & $Q^2$ & 0 & $Q$ \\
 $R^1_c$ & 1 & $Q^2$ & $Q^2$ & 0 & $Q^6$ & 0 & $Q$ \\
 $R^1_e$ & 0 & $Q$ & 0 & $Q$ & $Q^6$ & $Q^6$ & 0 \\
\hline
 $R^2_a$ & $Q^2$ & $Q^4$ & $Q^4$ & 0 & $Q^8$ & 0 & $Q^4$ \\
 $R^2_b$ & $Q$ & $Q^3$ & $Q^3$ & 0 & $Q^8$ & 0 & $Q^3$ \\
 $R^2_d$ & 1 & $Q^2$ & $Q^2$ & 0 & $Q^8$ & 0 & $Q^2$ \\
 $R^2_e$ & 1 & $Q^2$ & $Q^2$ & 1 & $Q^8$ & 0 & $Q^2$ \\
\end{tabular}
\end{table}

As we can see from Table~\ref{tab.01}, 
the high $Q$ behavior of each helicity amplitude 
in general depends on the reference frame. This is so because the 
helicities and the components of the current do mix in general,
although the physical form factors are of course identical 
for any $Q$ regardless of the reference frame. Only in frames
connected by a kinematic transformation that keeps the 
light-front time $\tau= t+z/c (= 0)$ invariant, 
the helicity amplitudes $G^+_{h'h}$ are the same\cite{JiMitchell}. 
Indeed, our results summarized in 
Table~\ref{tab.01} are essentially identical 
in kinematically connected frames such as DYW, 
Breit($\theta = \pi/2$) and TRF($\theta = \theta_0$)\footnote
{The reason for an extra power $Q$ for the Breit($\theta = \pi/2$)
and TRF($\theta = \theta_0$) in comparison to DYW can be understood 
by the kinematic factors in the relation between $(G_C,G_M,G_Q)$
and $G^+_{h^\prime h}$.}. Note that $\theta_0 \to \pi$ in the 
limit $Q \to \infty$. It is interesting to find that in all cases the behavior
of the helicity amplitudes in these frames is
consistent with the perturbative QCD predictions obtained in the $q^+=0$ frame.
Indeed, PQCD predicts\cite{LepageBrodsky} that the helicity-zero to
helicity zero matrix element $G^+_{00}$(or $G_d$) is the dominant
helicity amplitude at large $Q^2$\cite{BrodskyHiller}.
For example, in the deuteron form 
factor\cite{CarlsonHillerHolt} calculation using the factorization 
theorem of PQCD one can show
that the five intermediate gluons connecting the six quarks
can be arranged in such a way that the gluon polarizations
and quark helicitys alternate to allow a maximum amplitude
when the initial helicity zero state transits to the final
helicity zero state. Further, in the $q^+ = 0$ frame, PQCD predicts  
that the helicity-flip
amplitudes $G^+_{+0}$($G_c$) and $G^+_{+-}$($G_e$) are suppressed
by factors of $Q^{_1}$ and $Q^{_2}$, respectively
\begin{equation}
 G_c = a \frac{\Lambda_{QCD}}{Q} G_d \quad
 G_e = b \left(\frac{\Lambda_{QCD}}{Q}\right)^2 G_d
 \label{eq.5.50}
\end{equation}
where $a$ and $b$ are constants and there are also corrections of order
$\Lambda_{QCD}/m$ \cite{BrodskyHiller,CarlsonGross}.  Our results, based
on the naturalness condition, coincide with these PQCD predictions. From
the table, we also find that $G^+_{++}$($G_a$) should be suppressed by
two powers of $Q$ compare to the dominant $G^+_{00}$ in the high $Q$
limit. However, neither our analysis nor PQCD can fix the constants $a$
and $b$.  Both angular conditions, AC 1 and AC 2, are satisfied
independent from $a$ and $b$.  Thus, both angular conditions are
consistent with the PQCD predictions.

On the other hand, in the frames that are not connected to DYW by a
kinematical transformation the results are not consistent with the PQCD
predictions as one can see from Table~\ref{tab.01}.  Since there are
contributions from embedded states\cite{BakkerJi} in $q^+ \neq 0$
frames, there are no reasons why they should be consistent with the
leading-order PQCD predictions. Nevertheless, it is interesting to note that 
$G_d$ dominates regardless of the reference frame.  We now discuss some
details of AC 1 and AC 2 in each reference frame.

\subsubsection{Drell-Yan-West Frame}
\label{sec.V.A.1}
The first angular condition, AC 1, is simple. It reads
\begin{equation}
 e^{-i\phi} G_b + e^{i\phi} G_c = 0 .
 \label{eq.5.60}
\end{equation}
The leading $Q$-behavior of the l.h.s. of AC 1 is
\begin{equation}
 \frac{m}{Q} (R^1_b  G_b + R^1_c G_c)
 \stackrel{Q \to \infty}{\sim} 
-\frac{p^+}{2\surd 2} \, \left( 4 F_1 + 2 F_2 + \frac{m^2}{Q^2} F_3 \right)
+\frac{p^+}{2\surd 2} \, \left( 4 F_1 + 2 F_2 + \frac{m^2}{Q^2} F_3 \right) .
 \label{eq.5.70}
\end{equation} 
So, if we assume $F_3 \stackrel{Q \to \infty}{\sim} \frac{Q^2}{m^2} H_3$
and $F_1, F_2$ and $H_3$ are of the same order in $Q^2$ for $Q \to \infty$, then
both terms are equal in magnitude.

AC 2 is more involved, but still easy. Its l.h.s. behaves for $Q \to \infty$ 
to leading order as follows
\begin{eqnarray}
&&  \frac{m^2}{Q^2} (R^2_a G_a + R^2_b  G_b + R^2_d G_d + R^2_e G_e)
 \stackrel{Q \to \infty}{\sim}
 \nonumber \\
 && \quad {m^2 p^+}\left[ \frac{1}{2} (4F_1 +H_3) - (4 F_1 +2 F_2 +H_3)
+ \frac{1}{2} (4 F_1 + 4 F_2 + H_3) \right]. 
 \label{eq.5.80}
\end{eqnarray}

The term involving $G_e$ does not contribute in leading order. 

\subsubsection{Breit Frame}
\label{sec.V.A.2}
First AC 1. We multiply with $(m/Q)^4$
\begin{eqnarray}
 && \frac{m^4}{Q^4} (R^1_a G_a + R^1_b G_b + R^1_c G_c + R^1_e G_e ) 
 \stackrel{Q \to \infty}{\sim}
 \nonumber \\
 && \quad m^3\left[-\frac{1}{4}(4F_1+H_3)\sin^2\theta \cos\theta \right.
\nonumber \\
 && \quad\quad \quad  \left. -\frac{(4F_1 + 4F_2 +H_3)\cos\theta + \{4F_1 + F_2 (3+\cos 2 
\theta) + H_3\}} {8(1+\cos\theta)^2} \sin^4\theta \right. \nonumber \\
 && \quad\quad\quad   -\left. \frac{(4F_1 + 4F_2 +H_3)\cos\theta - \{4F_1 + F_2 (3+\cos\theta) + 
H_3\}} {8(1-\cos\theta)^2} \sin^4\theta \;\right].
 \label{eq.5.90}
\end{eqnarray}
Actually, $R_e^1 G_e$ is two orders $Q/m$ down compared to the 
other three terms. The contributions of three terms that remain in leading
order will depend on the angle $\theta$. For example, for $\theta = 0$ 
all vanish identically and we find that then the leading order is lower 
than $(Q/m)^4$. For $\theta=\pi/2$ only the terms $R_b^1 G_b$ and $R_c^1 G_c$
survive and cancel each other.

The leading order of AC 2 is $(Q/m)^5$. We find
\begin{eqnarray}
 & &  \frac{m^5}{Q^5} (R^2_a G_a + R^2_b G_b + R^2_d G_d + R^2_e G_e )
 \stackrel{Q \to \infty}{\sim} 
 \nonumber \\
 & & \quad m^3\left[ - \frac{4F_1 + H_3}{8\sqrt{2}}\sin^4\theta 
 + \frac{ 4F_1 + 
 2F_2(1+\cos\theta)+H_3}{4\sqrt{2}(1-\cos\theta)^2(1+\cos\theta)}\sin^6\theta
\right.  \nonumber \\
 & & \quad \quad \left. \quad
- \frac{(4F_1+4F_2+H_3)(3-4\cos\theta+\cos2\theta)}{16\sqrt{2}(1-\cos\theta)^4}
 \sin^6\theta \; \right], 
 \label{eq.5.100}
\end{eqnarray}
and again the term with $G_e$ is not of leading order. For $\theta \to 0$,
the first term is of order $\theta^4$ while the two others are of order 
$\theta^2$ and cancel each other exactly at this order. So, for small 
$\theta$ the contributions of $G_b$ and $G_d$ dominate. For $\theta=\pi/2$,
all three terms are of the same order. This situation corresponds exactly 
with AC 2 in the DYW frame.

\subsubsection{Target Rest Frame}
\label{sec.V.A.3}
Since the leading term in AC 1 is of order $(Q/m)^8$, we multiply it with 
$(m/Q)^8$ and find 
\begin{eqnarray}
 && \frac{m^8}{Q^8} (R^1_a G_a + R^1_b G_b + R^1_c G_c + R^1_e G_e )
 \stackrel{Q \to \infty}{\sim}
 \nonumber \\
 && \quad \frac{m^4 \sin^2\theta}{512\sqrt{2}}\left[\;
(-48-64\cos\theta-16\cos2\theta)F_1 \right.
\nonumber \\
 && \quad\quad\quad\quad\quad
 + \left. (-2 + 2 \cos2\theta -2\cos\theta + 2\cos\theta \cos2\theta)H_3
\right.\nonumber \\
 && \quad\quad\quad\quad\quad
 +\left. (48+64\cos\theta+16\cos2\theta)F_1
\right.\nonumber \\
 && \quad\quad\quad\quad\quad
 +\left. (-12\cos\theta -4\cos\theta \cos2\theta -16\cos^2\theta)H_3
\right.\nonumber \\
 && \quad\quad\quad\quad\quad
 +\left. (10 + 15 \cos\theta +6\cos2\theta + \cos3\theta)H_3 \; \right].
 \label{eq.5.110}
\end{eqnarray}
The contribution from $G_e$ is not of leading order. The other three
terms are comparable in size, but the details depend on the angle $\theta$. 

AC 2 is different, as only $G_b$ and $G_d$ contribute in leading order, 
which is $(Q/m)^{12}$. We find
\begin{eqnarray}
 & & \frac{m^{12}}{Q^{12}} (R^2_b G_b + R^2_d G_d)
 \stackrel{Q \to \infty}{\sim}
 \nonumber \\
 & & \frac{m^5 \sin^2\theta}{256\sqrt{2}}\left[-(4F_1 + 4F_2 +H_3) +
 (4F_1 + 4F_2 +H_3)\right] \nonumber \\
& & \times \cos\theta(1+\cos\theta)(3+4\cos\theta+\cos2\theta).
 \label{eq.5.120}
\end{eqnarray}

\section{Conclusions}
\label{sec.VI}
In this work we elaborated the frame dependence of the angular
conditions for spin-1 systems. We found that there is an additional
angular condition besides the well-known one given by
Eq.~(\ref{eq.1.40}).  In the $q^+ = 0$ frame including DYW,
Breit($\theta = \pi/2$) and TRF($\theta = \theta_0$), we find that the
additional condition is very simple involving only two helicity
amplitudes as shown in Eq.~(\ref{eq.4.40}) and most quark models rather
easily satisfy it. Thus, it doesn't seem to provide as strong a
constraint as the usual condition Eq.~(\ref{eq.4.50}).  However, in
$q^+ \neq 0$ frames, the additional condition is generally as
complicated as the usual one. Since the $q^+ = 0$ frame ({\it e.g.}
DYW) is in principle restricted to the spacelike region of the form
factors, it may be useful to impose the additional condition in
processes involving the timelike region. Nevertheless, it seems rather
clear from our spin-1 form factor discussion that the analysis of
exclusive processes is greatly simplified in the DYW frame and in
general $q^+ = 0$ frames. We note that the angular conditions given by
Eqs.~(\ref{eq.4.40}) and (\ref{eq.4.50}) are identical in any frame
connected to the DYW frame by kinematical transformations.

We also find that both angular conditions in the $q^+ = 0$ frame are
consistent with the PQCD predictions. Our predictions for the
$Q$-dependence of the helicity amplitudes based on the naturalness
condition as well as the angular condition are also consistent with the
PQCD predictions given by Eq.~(\ref{eq.5.50}). However, the
proportionality constants $a$ and $b$ can be fixed neither by our
analysis nor by PQCD. Some other inputs such as experimental data are
needed to find these values. For example, in the deuteron analysis a
value near 5 was obtained for $a$\cite{Kobushkin}.  Nevertheless, it is
interesting to note that for some particular values of $a$ and $b$ the
relations among $F_1, F_2$ and $F_3$ are greatly simplified. For
$a=b=0$, we find that $F_2 /F_1 = -2$ and $F_3/F_1 =0$, which are
identical to Eq.~(\ref{eq.5.40}) for a point particle.  Since the form
factors for a point particle do not depend on $Q^2$ at tree level, one
can understand this universality result rather easily. Also, for $a =
\sqrt{2} m/\Lambda_{\rm QCD}$ and $b = 2 m^2/\Lambda^2_{\rm QCD}$ we
find that $F_2/F_1 = -1$ and $F_3/F_1 = -1$. Even though the results
are simple for these particular values of $a$ and $b$, it is not yet
clear what their importance is. In order to analyze the values of $a$
and $b$, one may need to have some sort of bound-state information for
the spin-1 system. Work along this line, using a simple but exactly 
solvable mode, is in progress.

\acknowledgements
This work was supported in part by a grant from the U.S. Department of 
Energy (DE-FG02-96ER 40947) and the National Science Foundation 
(INT-9906384). 
BLGB wants to thank the Department of Physics at NCSU for their hospitality
when this work was completed. CRJ thanks for the hospitality of the Physics
Department at Vrije Universiteit during his visit.
The North Carolina Supercomputing Center and the National 
Energy Research Scientific Computer Center are also acknowledged for the 
grant of computing time. 

\appendix
\section{Lorentz Transformations}
\label{sec.A}
Most of the formulas given here can be found in or are based on the paper
by Leutwyler and Stern\cite{LS}(See also a recent 
literature\cite{JMrecent}.).

\subsection{Instant Form}
\label{sec.A.1}
If we write $E=\sqrt{\vec{p}^{\, 2} + m^2}$, then a pure boost from the rest 
frame to frame where the four momentum is $(E, \vec{p})$ is given by
\begin{equation}
 L^\mu_{\; \nu} =
 \left(
 \begin{array}{rrrr}
 \frac{E}{m} & \frac{p_x}{m} & \frac{p_y}{m} & \frac{p_z}{m} \\
 \frac{p_x}{m} & 1+ \frac{p^2_x}{m(E+m)} & \frac{p_x p_y}{m(E+m)} &
 \frac{p_x p_z}{m(E+m)} \\
 \frac{p_y}{m} & \frac{p_y p_x}{m(E+m)} & 1+\frac{p^2_y}{m(E+m)} &
 \frac{p_y p_z}{m(E+m)} \\
 \frac{p_z}{m} & \frac{p_z p_x}{m(E+m)} & \frac{p_z p_y}{m(E+m)} &
 1+\frac{p^2_z}{m(E+m)}
 \end{array}
 \right).
 \label{eq.A.10}
\end{equation}

\subsection{Front Form}
\label{sec.A.2}
In order to facilitate the derivation we define the connection between
the usual four-vector components $p^\mu = (p^0, p^1, p^2, p^3)$ and the 
front-form components $p^\mu_{\rm ff} = (p^+, p^1, p^2, p^-)$ with the 
definition
\begin{equation}
 p^\pm = \frac{p^0\pm p^3}{\surd 2},
 \label{eq.A.20}
\end{equation}
which can be written with the help of the matrix
\begin{equation}
 \eta = \left(
 \begin{array}{cccc}
 \frac{1}{\surd 2} & 0 & 0 & \frac{1}{\surd 2} \\
 0 & 1 & 0 & 0 \\
 0 & 0 & 1 & 0 \\
 \frac{1}{\surd 2} & 0 & 0 & -\frac{1}{\surd 2}
 \end{array}
 \right).
 \label{eq.A.30}
\end{equation}
Using $\eta$ we can write the relations Eq.~(\ref{eq.A.20}) as
\begin{equation}
 p^\mu_{\rm ff} = \eta^\mu_{\; \nu} p^\nu .
 \label{eq.A.40}
\end{equation}
The matrix $\eta$ has the nice property that it is idempotent. We can
use it to define the components of any tensor. As an example we
transform the metric tensor $g$ to $g_{\rm ff}$
\begin{equation}
 g_{\rm ff} = \eta \, g \, \eta =
 \left(
 \begin{array}{cccc}
 0 & 0  & 0  & 1 \\
 0 & -1 & 0  & 0 \\
 0 & 0  & -1 & 0 \\
 1 & 0  & 0  & 0
 \end{array}
 \right).
 \label{eq.A.50}
\end{equation}
We could write the pure boost in front-form coordinates, if we wanted to, but we
don't, because we want to use the kinematical front-form boost, which we write
next.

The kinematical front-form boost is given by
\begin{equation}
 L_{\rm ff} (\vec{v}_\perp ;\chi) =
 \exp(-i \surd 2 \, \vec{v}^\perp \cdot \vec{E}^\perp ) \; \exp(-i\chi K^3),
 \label{eq.A.60}
\end{equation}
where $K^3 = M^{-+}$ is the third component of the boost generator
and the generators $E^1$ and $E^2$ are given by
\begin{equation}
 E^1 = M^{+1} = \frac{1}{\surd 2}\, (K^1 + J^2), \quad
 E^2 = M^{+2} = \frac{1}{\surd 2}\, (K^2 - J^1).
 \label{eq.A.70}
\end{equation}

By taking  specific values for the transverse velocity $\vec{v}^\perp$
and the hyperbolic angle $\chi$ we obtain the front-form boost from the
rest system where the momentum has components $p^\mu = (m/\surd 2, 0,
0, m/\surd 2)$ to the frame where it has components $(p^+, p^1, p^2,
p^-)$. We must account for the fact that the dispersion relation in the
front form is
\begin{equation}
 p^- = \frac{\vec{p}^{\perp \, 2} + m^2}{2 p^+},
 \label{eq.A.80}
\end{equation}
where we introduced the obvious notation
\begin{equation}
 \vec{p}^{\,\perp} = (p^1, p^2).
 \label{eq.A.90}
\end{equation}
The connection we need is
\begin{equation}
 e^\chi = \surd 2 \, p^+ /m, \quad
 \vec{v}^\perp = \vec{p}^{\,\perp} /(\surd 2 \, p^+).
 \label{eq.A.100}
\end{equation} 
The generators $E^1$ and $E^2$ are nilpotent and $K^3$ has also a simple form, 
viz
\begin{equation}
 K^3_{\rm ff} = i\, \left( \begin{array}{cccc}
 1 & 0& 0& 0 \\
 0 & 0& 0& 0 \\
 0 & 0& 0& 0 \\
 0 & 0& 0& -1
 \end{array}
 \right) , \quad
 E^1_{\rm ff} = \left( \begin{array}{cccc}
 0 & 0& 0& 0 \\
 1 & 0& 0& 0 \\
 0 & 0& 0& 0 \\
 0 & 1& 0& 0
 \end{array}
 \right) , \quad
 E^2_{\rm ff} = \left( \begin{array}{cccc}
 0 & 0& 0& 0 \\
 0 & 0& 0& 0 \\
 1 & 0& 0& 0 \\
 0 & 0& 1& 0
 \end{array}
 \right) .
 \label{eq.A.110}
\end{equation}
Thus, the explicit form for the boost is not difficult to determine. It is
\begin{equation}
 L_{\rm ff} (\vec{v}_\perp ;\chi) =
 \left( 
 \begin{array}{cccc}
 \surd 2 p^+/m& 0& 0& 0 \\
 \surd 2 p^1/m& 1& 0& 0 \\
 \surd 2 p^2/m& 0& 1& 0 \\
 \vec{p}^{\perp \, 2}/(m \surd 2 \, p^+)& p^1/p^+& p^2/p^+& m /(\surd 2 \, p^+ )
 \end{array}
 \right) .
 \label{eq.A.120}
\end{equation}
Indeed, if we act with $L_{\rm ff}$ on $(m/\surd 2, 0, 0, m/\surd 2)$ we find
for $p^\mu$: $(p^+, p^1, p^2, p^-)$ with $p^-$ given by Eq.~(\ref{eq.A.80}).
To this simple check we can add the defining property of the Lorentz 
transformations $L$, {\it i.e.}
\begin{equation}
 g^{\mu\nu} = L^\mu_{\;\alpha} L^\nu_{\; \beta} g^{\alpha\beta}, 
 \label{eq.A.130}
\end{equation}
which can be translated into matrix form as follows
\begin{equation}
 g = L^\top g L.
 \label{eq.A.140}
\end{equation}
This relation can be interpreted as a quasi-orthogonality condition on the
rows of the transformation symbol $L$. Needless to say that
the orthogonality condition must be implemented with the right form of the
metric $g$. 

\subsection{Helicity}
\label{sec.A.3}
The operator 
\begin{equation}
 L_{\rm ff} (0;\chi) = \exp(-i\chi K^3) =
 \left( 
 \begin{array}{cccc}
 \surd 2 p^+/m& 0& 0& 0 \\
 0 & 1& 0& 0 \\
 0 & 0& 1& 0 \\
 0 & 0 & 0 & m /(\surd 2 \, p^+ )
 \end{array}
 \right)
 \label{eq.A.150}
\end{equation}
commutes with $J^3$ because $K^3$ does. Therefore, the polarization vectors
\begin{equation}
 L_{\rm ff} (0;\chi) \, \epsnul(h) 
 \label{eq.A.160}
\end{equation}
are eigenvectors of $J^3$. If we next apply the front-form combinations of
rotations and boosts  
 $ \exp(-i \surd 2 \, \vec{v}^\perp \cdot \vec{E}^\perp ) $
to move the vector $(p^+, 0,0, m^2/(2 p^+))$ to $(p^+, p_x, p_y, (\vec{p}^{\perp \, 2} + m^2)/(2 p^+))$ and use the full LF boost Eq.~(\ref{eq.A.120}) to
obtain the boosted polarization vectors, we see that we can use the operator
\begin{equation}
 h_{\rm ff} = \exp(-i \surd 2 \, \vec{v}^\perp \cdot \vec{E}^\perp ) J^3 
 \exp(i \surd 2 \, \vec{v}^\perp \cdot \vec{E}^\perp ) .
 \label{eq.A.170}
\end{equation}
This operator, which we call the {\em LF helicity}, has the eigenvectors
$\varepsilon_{\rm ff} (h)$ with $h=0, \pm 1$, Eq.~(\ref{eq.2.140}),
and a fourth eigenvector $(0,0,0,1)$. The
latter does not correspond to a genuine polarization vector. It has only a
minus component, which means that it is orthogonal to all four vectors with
$p^- =0$, {\it i.e.} $p^+ \to \infty$.

The explicit form of $h_{\rm ff}$ is
\begin{equation}
 h_{\rm ff} =  i\, \left( \begin{array}{cccc}
 0& 0& 0& 0 \\ p^2/p^+& 0& -1& 0 \\ - p^1/p^+& 1& 0& 0 \\ 0 & p^2/p^+& -p^1/p^+& 0
 \end{array}
 \right) .
 \label{eq.A.180}
\end{equation}
One can write it in operator form as
\begin{equation}
 h_{\rm ff} = \frac{W^+}{P^+} = J^3 - \frac{P^1 E^2 - P^2 E^1}{P^+}.
 \label{eq.A.190}
\end{equation}
This operator is clearly a kinematic one, as $J^3$, $P^1$, $P^2$, $E^1$, and
$E^2$ all belong to the stability group of $x^+ =0$.

\section{Symmetries of frames and relations between different frames}
\label{sec.B2}
In this section we give the kinematical Lorentz transformations that
connect the different frames in specific cases. We stress that the
frames can be transformed into each other by general Lorentz transformation,
but only in special cases can this be done using elements from the
kinematical subgroup alone. 

The kinematical group is generated by $J^3$, $K^3$ and $E^1$ and
$E^2$.  As all frames are invariant under rotations about the $z$-axis,
we shall not discuss $J^3$. We can use this kinematical rotation to
remove the $\phi$-dependence of the angular conditions.  The
interesting transformations are $L_{\rm ff}(0; \chi)$ and 
$L_{\rm ff}(\vec{v}_\perp;0)$.

\subsection{Symmetries of frames}
\label{sec.B201}

\subsubsection{Boosts along the $z$-axis}
\label{sec.B21}
$L_{\rm ff}(0; \chi)$ is a symmetry of the Drell-Yan-West frame, but not of the
Breit frame or target rest frame.

\subsubsection{Transverse Boosts}
\label{sec.B22}
We write $L_{\rm ff}(\vec{v}_\perp;0)$ explicitly
\begin{equation}
 L_{\rm ff}(\vec{v}_\perp;0) = \left(
 \begin{array}{ccc}
 1 & 0 & 0 \\
 \surd 2 \, \vec{v}_\perp & 1 & 0 \\
 \vec{v}^{\, 2}_\perp & \surd 2 \, \vec{v}_\perp & 1
 \end{array}
 \right) .
 \label{eq.B.10}
\end{equation}
The transverse boosts are not symmetries of the target rest frame.

If we apply it to the DYW momentum transfer we find
\begin{equation}
 L_{\rm ff}(\vec{v}_\perp;0) q_{\rm DYW} =
 \left(0, Q\hat{n}, \frac{Q^2}{2p^+} + \surd 2 \, 
 Q \hat{n} \cdot \vec{v}_\perp \right).
 \label{eq.B.20}
\end{equation}
If one generalizes the definition of the DYW frame to $q^+ = 0$, then this 
transformation is a symmetry of this frame, if one allows for 
a perpendicular momentum in the initial state
\begin{equation}
 \vec{p}_\perp = \surd 2\, p^+ \vec{v}_\perp ,
 \label{eq.B.30}
\end{equation}
otherwise, insisting on $\vec{p}_\perp =0$ in the DYW frame, it is not.

In the Breit frame we find for the transformed momentum transfer
\begin{equation}
 L_{\rm ff}(\vec{v}_\perp;0) q_{\rm Breit} = (Q\cos\theta/\surd 2,
 Q(\sin\theta \hat{n} + \cos\theta \vec{v}_\perp),
 Q(-\cos\theta + 2\sin\theta  \hat{n} \cdot \vec{v}_\perp +
 \cos\theta \vec{v}^2_\perp)).
 \label{eq.B130}
\end{equation}
If we require this vector to have the form
\begin{equation}
 q'_{\rm Breit} = (Q\cos\theta/\surd 2, Q\sin\theta \hat{n}',
 -Q\cos\theta/\surd 2),
 \label{eq.B.50}
\end{equation}
then we must find a vector $\vec{v}_\perp$ that satisfies
\begin{equation}
 (\sin\theta \hat{n} + \cos\theta \vec{v}_\perp)^2 = \sin^2 \theta.
 \label{eq.B.60}
\end{equation}
There are two classes of solutions: either $\cos\theta = 0$ and $\hat{n} \cdot
\vec{v}_\perp = 0$, or $ \cos\theta \vec{v}^{\, 2}_\perp + 2 \sin\theta \hat{n} \cdot
\vec{v}_\perp = 0$. In the latter case the length of the velocity vector is
correlated with its direction through the relation
\begin{equation}
 v = - 2 \tan\theta \hat{n} \cdot \vec{v}_\perp .
 \label{eq.B.70}
\end{equation}
If we denote the azimuthal angles of $\hat{n}$ and $\vec{v}_\perp$ by $\phi$ 
and $\psi$ respectively then the vector $\hat{n}'$ in Eq.~(\ref{eq.B.50}) is 
given by %
\begin{equation}
 \hat{n}' = (-\cos(2\psi - \phi), - \sin(2\psi - \phi)).
 \label{eq.B.80}
\end{equation}
We conclude that there is a class of transverse boosts that leaves the Breit 
frame invariant.

\subsection{Relations between different frames}
\label{sec.B23}
If we want two reference frames to be connected by a Lorentz transformation,
we need to verify that both the initial momenta ($p$) and the momentum
transfers ($q$) are related by the same transformation. 

In the case of TRF and DYW the two are identical if $p^+ = m/\surd
2$ and in addition $\beta \, \sin \theta = 1$. The corresponding angle
we denote by $\theta_0$. The latter condition ensures that the momentum
transfer in the TRF has vanishing plus-component.  Clearly, for every
value of $Q$ there is an angle, $\theta_0$, for which the TRF and the
DYW are kinematically connected.

If we try the same for the TRF and the Breit frame, we find that they
are kinematically related for all Q at $\theta = 0$.

The DYW and the Breit frame can only be related for $\theta = \pi/2$. Then
the momentum transfer in the Breit frame has the form
\begin{equation}
 q_{\rm Breit} = ( 0, Q \hat{n}, 0).
 \label{eq.B.90}
\end{equation}
We now try to find the transformation that transforms the momentum transfer in
the DYW frame into this special vector. If we write $\vec{v}_\perp = v 
\hat{v}_\perp$, then we find the parameters
\begin{equation}
 \hat{v}_\perp = - \hat{n}, \quad
 v = \frac{Q}{2 m \beta}, \quad e^\chi = \frac{m \beta}{\surd 2 p^+}.
 \label{eq.B.100}
\end{equation}
We see that for any value of $Q$ we can connect the DYW frame to the
Breit frame with $\theta = \pi/2$.

The main conclusion from this exercise is that the three frames considered here
are only in special cases related by kinematical Lorentz transformations.
In these cases the angular conditions are the same. In all other cases we
find non-equivalent angular conditions.

\section{Polarization Tensors}
\label{sec.C}
\subsection{DYW}
\noindent $G^+(1)$
\begin{equation}
 a_1 = 2 p^+ , \quad
 b_1 = -\frac{\surd 2 p^+ Q}{m} e^{i \phi} , \;
 c_1 = \frac{\surd 2 p^+ Q}{m}e^{-i \phi} , \;
 d_1 = p^+ \left(2 - \frac{Q^2}{m^2} \right), \;
 e_1 = 0,
 \label{eq.C.10}
\end{equation}
$G^+(2)$
\begin{equation}
 a_2 = 0, \quad
 b_2 = -\frac{p^+ Q}{\surd 2 m}e^{i \phi} ,\quad
 c_2 = \frac{p^+ Q}{\surd 2 m}e^{-i \phi} ,\quad
 d_2 = - \frac{ p^+ Q^2}{m^2}, \quad
 e_2 = 0,
 \label{eq.C.20}
\end{equation}
$G^+(3)$
\[
 a_3 =  \frac{p^+ Q^2}{2m^2} , \;
 b_3 = -\frac{p^+ Q^3}{2 \surd 2 m^3} e^{i \phi} ,\;
 c_3 =  \frac{p^+ Q^3}{2 \surd 2 m^3} e^{-i \phi} ,\;
\]
\begin{equation}
 d_3 = -\frac{p^+ Q^4}{4m^4} ,\;
 e_3 = -\frac{p^+ Q^2}{2m^2} e^{2i \phi}.
 \label{eq.C.30}
\end{equation}

\subsection{Breit Frame}

\noindent $G^+(1)$
\begin{eqnarray}
 a_1 & = & \quad \surd 2 m \beta , \nonumber \\
 b_1 & = & -\frac{2m\beta^2 Q\sin\theta}{2m\beta - Q \cos \theta} \, e^{i \phi},
 \nonumber \\
 c_1 & = & \quad \frac{2m\beta^2 Q\sin\theta}
 {2m\beta + Q \cos \theta} \,e^{-i \phi} ,
 \nonumber \\
 d_1 & = & \quad \frac{\surd 2\beta m[Q^2 \cos^2 \theta + 2\beta^2 
 (2m^2 - Q^2 \sin^2 \theta)]}
 {4\beta^2 m^2 - Q^2 \cos^2 \theta},
 \nonumber \\
 e_1 & = & \quad 0.
 \label{eq.C.40}
\end{eqnarray}
$G^+(2)$
\begin{eqnarray}
 a_2 & = & \quad 0, \nonumber \\
 b_2 & = &
 - \frac{\beta Q(2m\beta + Q \cos \theta)\sin \theta}
 {2 (2m\beta - Q \cos \theta)} \, e^{i \phi},
 \nonumber \\
 c_2 & = & \quad
 \frac{\beta Q(2m\beta - Q \cos \theta)\sin \theta}
 {2 (2m\beta + Q \cos \theta)} \, e^{-i \phi},
 \nonumber \\
 d_2 & = & \quad
\frac{\surd 2 \beta mQ^2[1-\beta^2 + (1+\beta^2)\cos 2\theta]}
 {4\beta^2 m^2 - Q^2 \cos^2 \theta}
 \nonumber \\
 e_2 & = & \quad 0 .
 \label{eq.C.50}
\end{eqnarray}
$G^+(3)$
\begin{eqnarray}
 a_3 & = & \quad
 \frac{\surd 2 m \beta^3 Q^2 \sin^2 \theta} {4\beta^2 m^2 - Q^2 \cos^2 \theta},
 \nonumber \\
 b_3 & = & - \frac{2m\beta^3 Q^2 
 (2m \cos\theta + \beta Q \sin^2 \theta)\sin\theta}
 {(2m\beta - Q \cos \theta)^2 (2m\beta + Q \cos \theta)} \, e^{i \phi},
 \nonumber \\
 c_3  & = & - \frac{2m\beta^3 Q^2
 (2m \cos\theta - \beta Q \sin^2 \theta)\sin\theta}
 {(2m\beta + Q \cos \theta)^2 (2m\beta - Q \cos \theta)} \, e^{-i \phi},
 \nonumber \\
 d_3 & = & \frac{2\surd 2 m \beta^3 Q^2 
 (4m^2 \cos^2 \theta - \beta^2 Q^2 \sin^4 \theta)}
 {(4\beta^2 m^2 - Q^2 \cos^2 \theta)^2},
 \nonumber \\
 e_3 & = & - \frac{\surd 2 m\beta^3 Q^2 \sin^2\theta}
 {4\beta^2 m^2 - Q^2 \cos^2 \theta}\, e^{2i \phi} ,
 \label{eq.C.60}
\end{eqnarray}

\subsection{TRF}
\noindent $G(1)$
\begin{eqnarray}
 a_1 & = &  \frac{2m + \kappa + \beta Q \cos\theta}{\surd 2}, \nonumber \\
 \nonumber \\
  b_1 & = & -\frac{\beta Q (2m + \kappa + \beta Q \cos\theta) \sin \theta}
 {2m} \, e^{i \phi} 
 \nonumber \\
 c_1 & = &  \frac{\beta Q (2m + \kappa + \beta Q \cos\theta) \sin \theta}
 {2(m + \kappa + \beta Q \cos\theta)} \, e^{-i \phi} ,
 \nonumber \\
  d_1 & = & \frac{(2m + \kappa + \beta Q \cos\theta)
 [2m^2 + 2m\kappa + \kappa^2 + 2\beta (m + \kappa)Q\cos\theta +
 \beta^2 Q^2 \cos 2\theta]}
 {2\surd 2m (m + \kappa + \beta Q \cos\theta)},
 \nonumber \\
 e_1 & = & 0.
 \label{eq.C.70}
\end{eqnarray}
$G(2)$
\begin{eqnarray}
 a_2 & = &  0 \nonumber \\
 b_2 & = &   -\frac{\beta Q (m + \kappa + \beta Q \cos\theta) \sin \theta}
 {2m} \, e^{i \phi} ,
 \nonumber \\
 c_2 & = & \quad
  \frac{\beta m Q \sin\theta}{2(m + \kappa + \beta Q \cos\theta)} \,
 e^{-i \phi} ,
 \nonumber \\
 d_2 & = & 
 \frac{(2m + \kappa + \beta Q \cos\theta)
 (\kappa^2 + 2\beta \kappa Q\cos\theta + \beta^2 Q^2 \cos 2\theta)}
 {2\surd 2 m (m + \kappa + \beta Q \cos\theta)},
 \nonumber \\
 e_2 & = & 0. 
 \label{eq.C.80}
\end{eqnarray}
$G(3)$
\begin{eqnarray}
 a_3 & = & \quad \frac{\beta^2 Q^2
 (2m + \kappa + \beta Q \cos\theta) \sin^2 \theta}
 {4\surd 2 m (m+\kappa+\beta Q \cos\theta)} ,
 \nonumber \\
 b_3 & = & -\frac{\beta Q (2m + \kappa + \beta Q \cos\theta) 
 [\kappa^2 + 2m\kappa + \beta^2 Q^2 + 2 \beta (m + \kappa) Q \cos\theta]
 \sin\theta}
 {8m^2 (m+\kappa+\beta Q \cos\theta)} \, e^{i\phi} ,
 \nonumber \\
 c_3 & = & - \frac{\beta Q (2m + \kappa + \beta Q \cos\theta)
 [\kappa(2m + \kappa) + 2\beta (m + \kappa) Q \cos\theta
 + \beta^2 Q^2 \cos 2\theta]\sin\theta}
 {8 m (m+\kappa+\beta Q \cos\theta)^2} \, e^{-i\phi} ,
 \nonumber \\
 d_3 & = &\quad 
 \frac{(2m + \kappa + \beta Q \cos\theta)}
 {8 \surd 2 m^2 (m+\kappa+\beta Q \cos\theta)^2}
 [\kappa^2 + 2m\kappa + \beta^2 Q^2 + 2 \beta (m + \kappa) Q \cos\theta]
 \nonumber \\
 &&
 \times [\kappa(2m + \kappa) + 2\beta (m + \kappa) Q \cos\theta
 + \beta^2 Q^2 \cos 2\theta] ,
 \nonumber \\
 e_3 & = & -\frac{\beta^2 Q^2(2m + \kappa + \beta Q \cos\theta) \sin^2 \theta}
 {4 \surd 2 m (m+\kappa+\beta Q \cos\theta)} \, e^{2i \phi}.
 \label{eq.C.90}
\end{eqnarray}

\end{document}